\documentstyle[12pt,psfig]{article}
\renewcommand{\thepage}{\arabic{page}}
\newcommand{\nc}{\newcommand}
\nc{\beq}{\begin{equation}} \nc{\eeq}{\end{equation}}
\nc{\beqa}{\begin{eqnarray}} \nc{\eeqa}{\end{eqnarray}}
\nc{\lsim}{\begin{array}{c}\,\sim\vspace{-21pt}\\< \end{array}}
\nc{\gsim}{\begin{array}{c}\sim\vspace{-21pt}\\> \end{array}}

\newcommand{\drawsquare}[2]{\hbox{%
\rule{#2pt}{#1pt}\hskip-#2pt
\rule{#1pt}{#2pt}\hskip-#1pt
\rule[#1pt]{#1pt}{#2pt}}\rule[#1pt]{#2pt}{#2pt}\hskip-#2pt
\rule{#2pt}{#1pt}}

\newcommand{\Yfund}{\raisebox{-.5pt}{\drawsquare{6.5}{0.4}}}
\newcommand{\Yasymm}{\raisebox{-3.5pt}{\drawsquare{6.5}{0.4}}\hskip-6.9pt%
        \raisebox{3pt}{\drawsquare{6.5}{0.4}}}

\begin{document}

\begin{titlepage}

{\hbox to\hsize{hep-th/9803107 \hfill Fermilab-Pub-98/089-T}}
{\hbox to\hsize{March 1998 \hfill UCSD-PTH-98-08}}
\bigskip

\begin{center}

\vspace{.5cm}

\bigskip

\bigskip

\bigskip

{\Large \bf   Dynamical Supersymmetry Breaking}

\bigskip

\bigskip

\bigskip

\bigskip

{\bf Erich Poppitz}$^{\bf a}$ and {\bf Sandip P. Trivedi}$^{\bf b}$ \\

\smallskip

{\tt  epoppitz@ucsd.edu,  trivedi@fnal.gov}

\bigskip

$^{\bf a}${\small \it Department of Physics \\
University of California, San Diego\\
La Jolla, CA 92093-0319, USA}

\bigskip

$^{\bf b}${ \small \it Fermi National Accelerator Laboratory\\
  P.O. Box 500\\
 Batavia, IL 60510, USA\\}

\bigskip

\bigskip

{\small 
Submitted to {\em Annual 
Review of Nuclear and Particle Science}, vol. 48.}

\bigskip

\bigskip

{\bf Abstract}
\end{center}

Dynamical supersymmetry breaking is a fascinating theoretical problem.
It is also of  phenomenological significance. A better understanding of
this phenomenon can help in model building, which in turn is useful in
guiding the search for supersymmetry.
In this article, we review the recent developments in the field.
We  discuss a few examples, which allow us to illustrate the main ideas
in the subject. In the process, we also show how the techniques of holomorphy
and duality come into play.  Towards the end we indicate how
 these developments have helped
in the study of gauge mediated supersymmetry breaking.
The review is intended for someone with a prior knowledge of supersymmetry 
who wants to find out about the recent progress in  this field.

\end{titlepage}

\baselineskip=18pt

\renewcommand{\thepage}{\arabic{page}}

\setcounter{page}{1}

\section{Introduction.}

Supersymmetry is a beautiful idea in theoretical physics. 
Unlike any conventional symmetry, it relates bosons  and fermions. 
It has proved   important in   many 
of the  major theoretical developments in recent times. 
For example, it  plays  a vital role   in string theory. 

There are phenomenological   reasons that make supersymmetry 
attractive as well.   The standard model presents us with a puzzle:
why is the electroweak scale so much smaller than the Planck scale?
This puzzle
 is called the hierarchy problem.  
Supersymmetric theories   promise  to 
solve this problem. The Higgs particle  can be  naturally incorporated as a  light 
elementary scalar  in these theories. 
Quadratically  divergent contributions to  its  mass are  then automatically
canceled by equal and opposite 
contributions arising from fermions.  
Moreover,
in  supersymmetric  extensions  of the standard  model, the large  top Yukawa 
coupling,  together with radiative effects, provides  a mechanism to break 
electroweak symmetry. 

But these positive features come at a  price:  by pairing 
fermions with bosons,  supersymmetry 
doubles the number of known  particles. 
The extra particles must clearly be heavy, leading to the conclusion that
supersymmetry must be broken in nature. 

 We do not have a good understanding of  how  this  breaking of 
supersymmetry   might   happen. 
Theoretically, as 
we will see, this is a  fascinating and challenging  question.  
It is of phenomenological importance  as  well.    
The  phenomenology  of supersymmetric  extensions of the 
standard model depends  in
an important way  on the masses of the superpartners  and the other soft
parameters, which 
 are all ultimately determined by  how  supersymmetry breaks.   In the absence
of  a  better  understanding of supersymmetry   breaking, 
 the  soft parameters are
taken to be arbitrary, resulting in  a huge parameter space. This  makes  a
thorough exploration of the resulting  phenomenology daunting.  A better
understanding of  the mechanisms of supersymmetry  breaking can, in turn, 
help   in  exploring    scenarios  with   restricted choices of the soft
parameters.  Such explorations  are 
useful in guiding the experimental search 
for supersymmetry.

In this review, we will discuss  various  mechanisms for  
dynamical supersymmetry 
breaking. 
 In the supersymmetric context, the
electroweak scale is  ultimately related to the supersymmetry 
breaking scale. Thus,
the hierarchy  problem can be recast in the form: 
why is the supersymmetry breaking scale so much  smaller 
than the Planck scale? 
Dynamical supersymmetry breaking 
provides the most attractive
answer to this question \cite{wittendsb}.  The idea is  that 
non-perturbative effects in 
a gauge theory are responsible for supersymmetry breaking.  For 
these effects to be important,
 the gauge coupling must be large. Moreover,  asymptotic freedom tells us that 
this can happen at a scale much lower  than the Planck scale. 
Thus we have  an appealing answer to the hierarchy problem:
the electroweak scale is  so much 
lower  than the Planck scale because gauge 
couplings only  run logarithmically. 

Recently, there has been phenomenal progress 
in our understanding 
of the dynamical behavior of supersymmetric gauge theories  
\cite{S2}-\cite{Shifman}. 
This progress  in turn  has lead to a better understanding of 
dynamical supersymmetry
breaking \cite{ADS1}-\cite{LT97}.  
Our main aim is to review  some of these
developments.  Specifically, we will  study  theories with
 $N=1$  global supersymmetry in $4$ dimensions. 
The restriction 
to $N=1$ supersymmetry  arises because of the phenomenological
requirement of chiral matter content. 
Considering only globally supersymmetric
  theories is less well motivated.  
 We do so here, first,
 because the recent progress has mostly been confined to such
 theories, and, second, because it allows for constructing models where the
 supersymmetry-breaking dynamics takes place at 
scales  low enough to be observed in the 
 foreseeable future. The study of such models 
carries a certain phenomenological appeal.

A very large number of  models    exhibiting  dynamical supersymmetry 
breaking have been constructed,  using the  newly developed techniques, in the 
recent past \cite{ISS}-\cite{LT97}. 
Clearly,  it would be pointless to  try and describe them all.  Instead,
we will attempt to build up an understanding of  
  supersymmetry breaking by studying a few  illustrative examples. 
 The general progression will
be from 
simpler to more complicated theories.   As the reader will see,
many of the main ideas  will recur throughout this   study  in  different
contexts. Wherever possible, we will also  attempt to make contact with other 
examples studied in the literature  (for short
reviews on the subject, see \cite{WS}-\cite{T1}). 

The  review is structured
as follows.  In Section 2, we first provide 
a very quick overview of some of the recent
developments  in supersymmetric gauge theories.   The discussion is  by no 
means  complete and is 
    intended  more to remind  the reader about some
salient features, which will be important in the discussion of supersymmetry
breaking.  Section 3 is a brief digression,  in which we study a supersymmetric
quantum  mechanics 
problem with supersymmetry breaking. This provides a
convenient setting  in which to  introduce some of the important ideas.  
Thereafter we turn to field theories. 
First,  in Section 4,
 some general features
of supersymmetry breaking  as well as  some simple examples of tree level
supersymmetry  breaking are discussed. Section 5.1 
then deals with calculable
models of dynamical supersymmetry breaking. 
In these models the low-energy  effective theory  in which 
supersymmetry breaking   occurs  can be completely controlled.  
This allows  a  
great deal to be learned about the resulting supersymmetry breaking 
ground state.  Section 5.2
deals with more  
complicated  theories. In some of these (Sections 5.2.1-5.2.3), 
we will be able to 
definitely establish supersymmetry breaking without 
being able to calculate in detail where the resulting 
vacuum lies. In other instances (Section 5.2.4),
  we  rely on the global symmetries and  the 
Witten index to plausibly argue that supersymmetry breaking occurs. 
Finally, in Section 6, we describe   how  some of these
theories of dynamical supersymmetry 
breaking might apply to nature in the gauge-mediated supersymmetry 
breaking scenario.

\section{Key ideas in the study of nonperturbative 
              supersymmetric gauge  dynamics.}

As was mentioned in the Introduction,  recently there has  been a great deal 
of progress in our understanding  of the non-perturbative dynamics of 
supersymmetric gauge theories. 
In this  section, we  briefly  discuss  some of the key ideas  that have played
an important role in these developments. We then also review
the case of supersymmetric  QCD to illustrate the different kinds of
non-perturbative effects  that  can occur in  a gauge theory.  
For an in-depth discussion of supersymmetric gauge theory dynamics, 
we refer the reader to the reviews \cite{IS}, \cite{Peskin}, \cite{Shifman}.

The recent progress in our understanding of four dimensional  $N=1$ 
supersymmetric gauge theories was initiated by  the work of Seiberg 
\cite{S2}, \cite{S1}, \cite{S3} (for a review of 
important work on the subject in the 1980's, see \cite{AKMRV}). 
Two  central  ideas  have played a particularly important  role in these
developments:  
\begin{enumerate}
\item{{\it Holomorphy.} The key realization is
that the superpotential of the  Wilsonian effective action of supersymmetric 
theories is a holomorphic function of the chiral superfields.\footnote{Holomorphy
can,  in fact, be used to prove \cite{S2} the ``old" nonrenormalization theorem, 
 that the superpotential is not renormalized at any order of
perturbation theory \cite{nonrenorm}.}
In addition, one can regard the couplings of the theory (the strong 
coupling scale, $\Lambda$, or the various superpotential couplings) 
as expectation values of  nondynamical chiral superfields \cite{S2}. The
couplings can be assigned charges under various symmetries (which are broken 
by their expectation values). This leads to certain ``selection rules,"
restricting how the couplings can appear in the effective action.
Now the Wilsonian superpotential has to be a holomorphic function of the 
chiral superfields as well as the  couplings of the theory. 
Holomorphy in the fields and couplings, together with the requirement of
consistency with various limits and  the above mentioned 
``selection rules"  allow  one to exactly determine the Wilsonian 
superpotential of the theory in many cases \cite{S2}.}
\item{{\it Duality} is the second crucial ingredient in our understanding of 
$N=1$ supersymmetric theories. Generalizing the
notion of electric-magnetic duality in Maxwell electrodynamics, 
Seiberg suggested 
 that, in many cases, the infrared limit of a  supersymmetric gauge theory 
 (the ``electric" theory)  is equivalent to the infrared limit of 
 another supersymmetric gauge theory (the ``magnetic" theory) \cite{S3}. 
 In some cases,
both theories flow to a nontrivial infrared fixed point and a description 
 of the fixed point in terms of either theory is appropriate (it is said then 
 that both theories are in
 the ``conformal window"), although 
 only one of the descriptions may be weakly coupled.
 In other cases, there is only one description of the infrared 
 theory, in terms of the  either  the electric or magnetic  degrees of freedom 
 (in these cases the relevant theory is often infrared free). 
It was also shown  that often  the exact superpotential in an electric confining theory
can be calculated by  doing an instanton calculation in a  weakly coupled and
completely higgsed dual theory.  While there is no proof of 
duality in $N=1$ theories, Seiberg's conjecture has 
survived many nontrivial tests \cite{S3}, 
\cite{IS}, \cite{APS}, most recently from brane dynamics  \cite{GK}.}
\end{enumerate}

Let us illustrate what these insights teach us  by studying  an $N=1$
$SU(N_c)$ gauge theory with  $N_f$ flavors of quarks (supersymmetric 
``QCD"). By this we mean 
 $N_f$ chiral superfields,
which we  denote as  $Q^{\alpha}_i $, $i=1, ..N_f$
in the  $\Yfund$  representation and  $N_f$ fields 
 ${\bar Q}^i_{\alpha} $, 
$i=1, ..N_f$ in the  $\overline{\Yfund}$  representation.  It is useful to  study the behavior
of this theory   as $N_f$ is varied.  We start by  first considering the case
$N_f=N_c-1$.  Classically, the theory  has  a  D-term potential   which 
is  set  to its minimum when the fields  satisfy the conditions: 
\beq
\label{flatsun}
Q^{\dagger ~i} ~T^a ~Q_i ~
-~ {\bar Q}^{\dagger}_i ~( T^a )^* ~{\bar Q}^{i} ~= ~0 
\eeq
for each group generator $T^a$. 
These conditions do not  select a unique vacuum. Instead,  the 
potential  has a set of flat directions.   
A general result \cite{LT} says that the flat directions 
can be parametrized  by gauge invariant chiral superfields.  
In the present case, there are $N_f^2$ flat directions. These correspond to the 
 ``meson " gauge invariants $M^i_j \equiv {\bar Q}^i  \cdot Q_j$. 
Along  these  flat directions the $SU(N)$ gauge symmetry is, generically, 
completely  broken.  The  $SU(N)$ vector multiplets are heavy  and the 
low-energy dynamics  can be described in  an effective theory   containing
only the mesons $M^i_j$. 

We now turn to the quantum theory. 
A  non-renormalization theorem \cite{nonrenorm} 
states that  the flat directions
are  not lifted  at  any order in perturbation theory.  But they can be lifted
non-perturbatively.  In fact, in the present case with $N_f=N_c-1$,
such a superpotential is induced by instanton effects  in the  Wilsonian  effective
theory for the  mesons, $M^i_j$. It has the form: 
\beq
\label{wsp1}
W_{NP}~= ~C ~{\Lambda^{b_0} \over  {\rm det} M } 
\eeq
Here $\Lambda$ is the strong coupling scale of the gauge theory, and 
$b_0 = 3N_c -N_f$  is the coefficient  of the one loop beta function. 
We note that holomorphy and the various symmetries   dictate that   a 
 superpotential  can  only have  this  form. An explicit constrained instanton
calculation   then  shows that  its coefficient $C$
 is  indeed non-zero \cite{ADS1}.   
 Note that the
superpotential results in a potential 
 energy  that goes to zero as   some of mesons go to infinity. 
This is often referred to as  ``runaway" behavior and 
results in an unstable ground state. 

 Now let  us vary $N_f$. When $N_f < N_c-1$,  the moduli 
are still described by  the meson fields $M^i_j$  for  the appropriate
number of  flavors.   
Generically in moduli space a  $SU(N_c-N_f)$ group is 
left unbroken.  This group confines,
 giving rise at low energies  to  an  effective
theory  involving only the mesons. The 
non-perturbative  superpotential  in this effective theory 
  arises due to gaugino condensation  and  is  given by:
\beq
\label{wsptwo}
W_{NP}~=~ C_{N_c,N_f}~ \left( 
{ \Lambda^{b_0}  \over  {\rm det} M } \right)^{1 \over N_c-N_f}~.
\eeq

For $N_f>N_c-1$ things get more interesting.  For example, for 
$N_f=N_c$ one finds that  the flat directions include, 
besides the meson
chiral superfields, additional   ``baryons," $B \sim Q^{N_c}$  
and $ {\bar B} \sim {\bar Q}^{N_c} $.  
These are not all independent. Correspondingly,  the 
quantum theory has  a constraint relating them, 
which is implemented by adding a
term  in the superpotential:  
\beq
\label{wspthree}
W_{NP}~=~A~\left(~{\rm det} M ~-~B ~\bar{B}~ - ~\Lambda^{b_0}\right)~.
\eeq
Here $A$ is the Lagrange multiplier whose $F$ term implements the constraint. 
In the  quantum theory this moduli 
space is smooth  and  the  low-energy effective theory   in terms
of the mesons and baryons is valid everywhere in  moduli space.  In particular,
there  is no submanifold of the moduli space where extra  degrees of freedom
became massless. 
The theory with $N_f = N_c + 1$ also has a smooth quantum moduli space. 
The  main difference is that  the ``origin" is   part of the moduli space in the 
quantum theory  too. All the global symmetries are unbroken  at the 
origin and the  mesons and baryons satisfy  the 't Hooft anomaly matching 
there as well. 

Finally, for $N_f>N_c+1$ one finds that the theory has a dual  
``magnetic" description in terms 
of  an $SU(N_f-N_c)$ gauge theory with  $N_f$ flavors of quarks,
$q^i$ and $\bar{q}_i$, $i=1, ... N_f$. In addition, there  are  
$N_f^2$ chiral  superfields  $M^i_j$ which can be  identified with the 
mesons of the electric theory. Note that in the dual description  the mesons  
 are  elementary fields present in the  microscopic ``magnetic" theory. 
The dual theory also has a tree level superpotential of the form:
\beq
\label{wdual}
W~=~\bar{q}_i~M^i_j~ q^j~. 
\eeq

This brings us to the end of our  lightning review of  the recent developments. 
To summarize, holomorphy and duality help determine the 
appropriate low-energy degrees of freedom---the ones pertinent to   the 
 physics below the strong coupling
 scale of the gauge theory---and determine 
 the   exact superpotential  of 
 the low-energy effective  theory. 

How do these insights help study supersymmetry breaking?
We note that the 
 ideas described above  are  useful  in determining the low-energy
 dynamics of supersymmetric gauge theories.  
We  will encounter  many  instances,  
during the study of supersymmetry breaking,
in which, by adjusting an appropriate
coupling,   the scale  of   supersymmetry breaking  can be   made lower 
than   the  scale of   strong dynamics in the gauge theory. In these cases,
 the  breaking of supersymmetry  can be conveniently  studied  in a
low-energy  supersymmetric effective theory. The ideas described above will
prove very useful then   in  determining this  effective theory and studying 
its behavior. 

The effective theory in these cases  will  usually  only involve 
a set of chiral superfields, $\Phi^i$. 
 The  corresponding Wilsonian  effective lagrangian is  then given by a
supersymmetric nonlinear sigma model: 
\begin{equation}
\label{sigmamodel}
L_{eff} ~=~ \int d^4 \theta ~K( \Phi^\dagger, \Phi ) ~+~ \left( ~\int
d^2 \theta ~W( \Phi) ~+~ {\rm h.c.} \right) ~.
\end{equation}
Here $K$ is the K\" ahler 
potential of the low-energy theory (a real function of the chiral superfields
$\Phi^i$), and $W$ is the superpotential of the theory---a holomorphic function
of the chiral superfields (and couplings). The complete component expansion of
(\ref{sigmamodel}) can be found in \cite{WB}. Here we will only give the expression
for the scalar potential of the sigma model (\ref{sigmamodel}):
\begin{equation}
\label{potential}
V ~=~ W^*_{i^*} ~K^{-1 ~{i^*} j} ~W_j ~,
\end{equation}
 where $W_i = \partial W/\partial \Phi^i, W^*_j = \partial W^*/\partial 
 \Phi^{*~{i^*}}$,
 and
 $K^{-1 ~{i^*} j}$ is the matrix inverse to 
the K\" ahler metric 
$K_{i {j^*}} = \partial^2 K/\partial \Phi^i \partial \Phi^{*~{j^*}}$; 
in (\ref{potential}) all
 functions are understood to depend 
on the scalar components of the superfields only.

 As we will see in Sections 3 and 4,  
 supersymmetry is broken if and only if
 the vacuum energy is nonvanishing. 
Since the K\" ahler metric $K_{i {j^*}}$ (and its
 inverse) is a 
 positive definite matrix---so that $L_{eff}$ makes sense 
 as a physical theory---the
 potential\footnote{Strictly speaking, to find
  the ground state energy one has to use the 
1PI rather than the Wilsonian effective 
  action. The superpotentials
   of the 1PI and Wilsonian effective actions, however,  are identical
    because of the nonrenormalization theorem (for discussions see \cite{IS},
 \cite{Shifman}, \cite{PR1}).} 
$V$ from (\ref{potential})  is positive semi-definite.  It vanishes only if 
the $F$ trem conditions, 
 $W_i = \partial W/\partial \Phi^i  =  0$  are met for  all the fields $\phi_i$.
If, on the other hand, these $F$ term  conditions  cannot  all  be met, 
the  vacuum energy  must necessarily  not vanish and supersymmetry is
broken.   Once we have found,  by holomorphy and duality,
 the  correct degrees 
of freedom  and the exact superpotential
of the low-energy effective theory, 
we can say with certainty whether a given theory
breaks supersymmetry. 
 
 Upon inspection of the effective Lagrangian (\ref{sigmamodel}),
 one sees
 that other important physical properties of the low-energy theory, such as
 the expectation values of the  fields, the  
 vacuum energy, the masses and interactions of the light fields,
 depend on the K\" ahler potential. In some cases,
 we will be able to explicitly determine it, while 
in other cases we will be at least
able to establish that  the corresponding   K\" ahler metric is non-singular.

\section{A toy  model.}

In this section, we   digress from the study of field theories to 
 explain  some  key ideas of supersymmetry breaking in 
a quantum-mechanical setting.
We will consider in some detail Witten's supersymmetric quantum mechanics.
This example will be used to introduce several important concepts:
 the order parameter for supersymmetry breaking and the Witten index.
The hope is, that  by encountering them in a simpler context the
reader will gain a better appreciation for these ideas.

\subsection{Supersymmetry breaking in quantum mechanics.}

The quantum mechanical system  is that of a  
spin-1/2 particle moving
on the line \cite{wittendsb}. 
The state of the spin-1/2 particle is described by a 
two-component wave function 
$\Psi (x)$; the two components of $\Psi$   are the
wave functions of the particle with spin projections 
$+1/2$ and $-1/2$, respectively. 
The supersymmetric Hamiltonian  is:
\begin{equation}
\label{hamiltoniansusyqm}
H ~=~ {1\over 2}~p^2~+~{1\over 2}~
W^\prime (x)^2 ~+~{1\over 2}~
\sigma_3 ~W^{\prime \prime} (x)~.
\end{equation}
Here and below $\sigma_{1,2,3}$ denote the Pauli matrices, $W^\prime (x) = 
dW/dx$, etc.
The supersymmetry generators are:
\begin{equation}
Q_1 ~=~ {1\over 2}~\sigma_1 ~p ~+~{1\over 2}~\sigma_2~ W^\prime (x) ~, ~~
Q_2 ~=~ {1\over 2}~\sigma_2 ~p ~-~{1\over 2}~\sigma_1~ W^\prime (x) ~.
\label{qmsusygenerators}
\end{equation}
They obey the  supersymmetry algebra:
\begin{equation}
\label{qmsusyalgebra}
\{ ~Q_i ~,~ Q_j ~\} ~=~ \delta_{ij}~H ~,~ i, j = 1, 2, ~
\end{equation}
with $H$ given by (\ref{hamiltoniansusyqm}).
The function $W(x)$ is called the superpotential; it completely determines
the interactions (in order to fully underline the 
analogy with quantum field theory, we  have slightly changed 
notations from \cite{wittendsb}).
 Note that the Hamiltonian (\ref{hamiltoniansusyqm}) is 
 similar to the one  obtained in $3+1$ dimensional
renormalizable supersymmetric field theory with spin-0 
and spin-$1/2$ fields only:
all   
interactions are derived by the derivatives
of a single function, the superpotential $W(x)$. 
In the field theory case, the  
``spin-orbit" term corresponds to the Yukawa interaction between the 
bosons and fermions in the supermultiplet.   Also, we see from 
(\ref{qmsusyalgebra})  that  the supersymmetry generators
transform the $+1/2$  eigenstate of $\sigma_3$ to the one with eigenvalue $-1/2$.
Thus, 
these two eigenstates  are the analogue of bosons and  fermions in this 
quantum mechanics problem.  Note, in particular, that the Hamiltonian, 
eq.~(\ref{hamiltoniansusyqm}) 
commutes with  $\sigma_3$ and  does not change 
``fermion number". 

The first issue we want to discuss is the order parameter for supersymmetry
breaking. The spontaneous breaking of supersymmetry means that even 
though the dynamics is 
invariant under supersymmetry, the ground state is not. 
The noninvariance of the ground state $\vert 0 \rangle$ 
under supersymmetry transformations 
implies  that 
the supersymmetry generators $Q_i$ do not annihilate the ground state, 
$Q_i  \vert 0 \rangle \ne 0$. Consider now  the ground state 
energy of the system, $E_0$, and  the following chain of
equalities:
\begin{eqnarray}
\label{orderparameter}
E_0 ~\equiv ~\langle~ 0~\vert~ H ~\vert ~0~ \rangle 
=~2~\langle ~0~\vert~  Q_i~ Q_i~\vert ~0~ \rangle 
= ~2~\vert\vert ~ Q_i~  \vert~ 0~ \rangle ~\vert\vert^2 ~> ~0, ~{\rm iff} ~
 Q_i~  \vert~ 0~ \rangle \ne 0 ,
\end{eqnarray}
where we used the fact that the supersymmetry algebra (\ref{qmsusyalgebra})
relates the supersymmetric
 Hamiltonian to the square of the supersymmetry generators (there is no sum
over $i$ in eq.~(\ref{orderparameter})).
The inequality in (\ref{orderparameter})
 holds whenever supersymmetry is broken, 
i.e. $Q_i  \vert 0 \rangle \ne 0$. We thus see
that the ground state energy of a supersymmetric system is positive if and only
if supersymmetry is broken and zero if and only if
 supersymmetry is unbroken. 
The 
ground state energy is therefore
 the order parameter for supersymmetry breaking. 
We note that this conclusion 
trivially generalizes to quantum field theory: the relativistic
supersymmetry algebra reduces to (\ref{qmsusyalgebra}) in the rest 
frame of the system.

At the classical level---ignoring the spin-orbit interaction and
the zero-point energies---it is easy to see whether
supersymmetry is broken or not. 
We only have to look at the graph of the 
potential energy $V \sim W^{\prime^2}$. We have
shown three possibilities on Fig.~1. Fig.~1a shows a potential which is
everywhere positive. Thus, classically, the ground state energy is positive and
supersymmetry is broken. The potentials on Fig.~1b,c both allow for classical
states of zero energy, hence, classically, supersymmetry is unbroken. 

\begin{figure}[ht]
 \centerline{\psfig{file=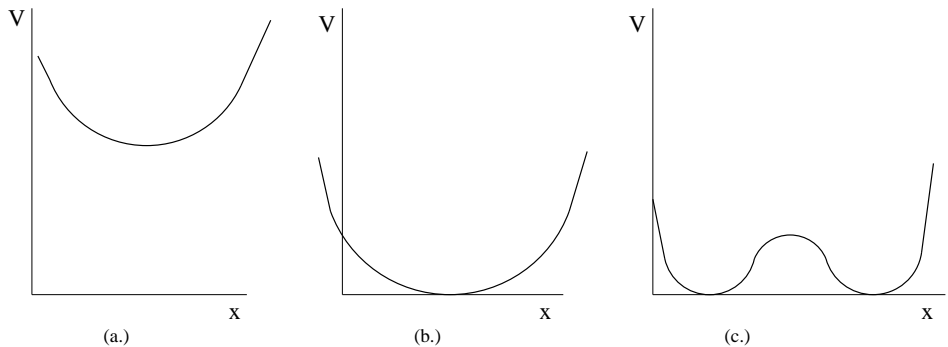}}
\caption{\protect \small The  three potentials discussed in the text: (a.) supersymmetry 
broken at tree level,
(b.) supersymmetry unbroken, (c.) supersymmetry unbroken at tree level, but
broken due to instantons (tunneling between the wells).}
\end{figure}

The classical 
approximation is, of course, not the whole story. 
It is natural to ask   whether quantum corrections can change the
classical answer. 
In   supersymmetric systems, as we will see throughout this article, 
it is often possible to give  exact answers to questions about 
the ground state. 
As discussed above, supersymmetry is unbroken if and only if 
 there is a normalizable zero energy state (we assume here
that the system has a discrete spectrum). 
Finding the zero-energy state implies solving 
the second order Schr\" odinger equation 
$H \vert 0 \rangle = 0$.
Now eq.~(\ref{orderparameter}) shows that $E_0 = 0$ if
 and only if $Q_i  \vert 0 \rangle = 0$; hence, it suffices, instead, 
 to solve the first order equation $Q_i  \vert 0 \rangle = 0$ (see the definition
of $Q_i$, eq.~(\ref{qmsusygenerators})). 
While a general second order equation 
can only be solved numerically, the corresponding first order equation
can be  solved for an arbitrary superpotential $W(x)$. 
Using simple Pauli matrix algebra, it is easy to check that 
\begin{equation}
\label{psizero}
\Psi_0 (x) = 
 e^{\sigma_3 ~W(x)} \left( \begin{array}{c}   c_1 \\  c_2 
 \end{array} \right)
= \left( \begin{array}{c} e^{W(x)}  c_1 \\
 e^{- W(x)} c_2 \end{array} \right)
\end{equation}
is the general solution for the zero-eigenvalue wavefunction.
$\Psi_0$ depends on two
integration constants $c_1, c_2$ and 
 is normalizable only in two cases: either 
$c_1= 0$ and $ W(x) \rightarrow ~ + \infty$ as $x
 \rightarrow \pm \infty$, or
$c_2 = 0$, while $W(x) \rightarrow - \infty$ as
 $x \rightarrow \pm \infty$. 
Thus a normalizable ground state of zero energy
exists only if the superpotential $W(x)$ is ``even at infinity," i.e. 
has the same limit ($+$ or $- \infty$) at both  $x = + \infty$ and $x = - \infty$. 
A smooth function $W(x)$ 
 with this property will necessarily have an odd number of extrema (and  
its derivative 
 $W^\prime$---an odd number of zeros).  Equivalently,  since 
$V(x) =  (W^\prime)^2/2$, we find 
 that the criterion for unbroken supersymmetry is that the 
 potential has an odd number of zeros. 

We can now revisit the three potentials on Fig.~1 and 
find whether supersymmetry is broken
or not  in the exact ground state. The potential
on Fig.~1a has no zeros, hence according to our criterion, 
supersymmetry, being broken at the tree level, 
remains broken once quantum corrections are included. 
The potential on Fig.~1b
has one minimum, hence  supersymmetry remains 
unbroken in the quantum theory.
Finally, in the case of Fig.~1c, the potential has an even number of zeros. 
Therefore, 
even though supersymmetry is unbroken at the classical level, it is broken by
quantum effects. We will see that all three cases have counterparts in quantum
field theory. 

It is the case depicted on Fig.~1c that will be of most interest for us. 
The  reason is that the breaking of supersymmetry in the 
supersymmetric system with a
double-well potential is due to nonperturbative effects---it occurs because
of tunneling between the two wells. 
We found earlier that in the classical approximation (and, even though we did not
show this, also in perturbation theory,
including the zero-point energy and the spin-orbit
interaction) the ground state energy vanishes and supersymmetry is unbroken.
The effect of tunneling can be evaluated in 
 the semiclassical approximation \cite{svh}.
The WKB formula for the ground state energy splitting
 gives  $E_0^{WKB} = \langle 0 \vert \hat{H} \vert 0 \rangle 
 \sim \hbar \omega  
\exp\left( -{1\over \hbar} \int d x \sqrt{2 V(x)} \right)  \ll \hbar \omega$
where $\omega$ is the  frequency of classical motion near the bottom of the 
well, and the integral is over the classically forbidden region of $x$.
Since, for appropriate parameters of the potential
(or,  in the  semiclassical $\hbar \rightarrow 
0$ limit),  the 
tunneling probability
 is exponentially 
 suppressed, the scale of supersymmetry breaking---the ground state 
energy---is  much smaller than the characteristic 
frequency of motion inside the wells. 

 The generation of
small scales, described above, also 
occurs naturally in many field theory models of 
dynamical supersymmetry breaking and 
is the key property 
that makes them   interesting candidates for explaining the hierarchy
of scales in nature.

\subsection{The Witten index.}

In the remainder of this section we will introduce another important
concept in the study of supersymmetric theories: the Witten index
\cite{wittenindex}. As we saw in our discussion of the supersymmetric 
quantum mechanical  model,  
whether supersymmetry is broken or not  depends
 only on the behavior of the superpotential $W(x)$ at
large $x$. Consequently, 
any continuous change of $W(x)$ that does not change its asymptotics 
at infinity will have no effect on whether the model breaks
supersymmetry. This is an indication that the issue of 
 supersymmetry breaking   has 
 topological nature: it depends only on 
asymptotics and global properties of the theory.
As a measure whether supersymmetry can break or not in a given model,
Witten introduced the index, ${\rm Tr} (-1)^F$, 
with $F$---the fermion number:
\begin{equation}
\label{wittenindex}
{\rm Tr}~(-1)^F ~\equiv~\sum_E ~ n_B (E) ~-~ n_F (E) ~=~ 
n_B (0) ~-~ n_F (0) ~.
\end{equation}
Here $n_{B (F)} (E)$ denotes the number of bosonic (fermionic) states of
energy $E$. The reason for the second equality is that in a supersymmetric
system every bosonic state of nonvanishing energy is
 degenerate with a  fermionic state (its superpartner), hence 
$n_B (E) = n_F (E)$ for $E \ne 0$, and only the zero energy states 
contribute to the index.\footnote{The pairing of nonzero energy states
is true whether or not supersymmetry is broken---in the case of broken 
supersymmetry, every state is degenerate with the state
obtained from it by adding a zero-momentum goldstino (Sect. 4.2).}
Note that since supersymmetry is unbroken if the theory has a
zero energy state, ${\rm Tr} (-1)^F \ne 0$ implies that 
the vacuum is supersymmetric. 

The main utility of the index (\ref{wittenindex}) is that 
it is invariant under changes of the Hamiltonian that do not change the
asymptotics of the potential (i.e. changing the Hamiltonian such
 that the added terms do not grow faster at infinity 
than the ones already present). 
This is because under continuous
changes of the parameters states can leave or descend to the zero energy
 level, but can do so only in pairs (because of the doubling of all $E\ne 0$
 levels), 
and hence do not affect the index.
Because of this invariance, a calculation of 
the index   at weak
coupling (i.e. in perturbation theory)  allows one to deduce
information about the ground state even at strong coupling.

We note that if a calculation of the index yields 
${\rm Tr} (-1)^F = 0$, without separate
knowledge of whether zero energy states exist 
($n_B(0) = n_F(0) \ne 0$)
one can not decide whether supersymmetry 
is broken.  In the case of vanishing index, under continuous 
deformations of the parameters of the 
Hamiltonian, all states can leave zero energy, so it 
is possible that supersymmetry is broken for some values of the 
parameters and not for others. In this case, 
more dynamical information is required to find
whether the ground state is supersymmetric.

As an example of the application of the Witten index, we can quickly 
calculate it in our quantum mechanical model
(we define $F = 1$ for spin projection $|+1/2 \rangle$ and $F=0$  for 
$|-1/2  \rangle$ ), for the three potentials of Fig.~1. In perturbation
theory, the potential of Fig.~1a  does not allow for any zero-energy states, hence the
index vanishes (and, as eq.~(\ref{psizero}) shows, 
there are no  states of zero energy in the exact solution, 
so supersymmetry is broken). 
The potential of Fig.~1b allows for a single zero energy state
(in the harmonic approximation near the minimum, depending on the sign
of $W^{\prime \prime}$, it is either bosonic or fermionic), hence 
${\rm Tr} (-1)^F =  \pm1$ (say), and supersymmetry is unbroken, even when 
all quantum effects are taken into account. Finally, Fig.~1c has two perturbative
zero-energy states---in the harmonic approximation 
to eq.~(\ref{hamiltoniansusyqm}) near each of the minima, one
of them has spin $+1/2$, and the other $-1/2$---so 
${\rm Tr} (-1)^F = 0$ and supersymmetry can be broken (and, as
the exact solution, eq.~(\ref{psizero}) shows, indeed is).

We should also  note that  by continuously changing 
 the parameters we can not
 interpolate between the theories of Fig.~1a and Fig.~1b (or Fig.~1b
and Fig.~1c) without changing
the Witten index. In order to 
deform, say, the potential of Fig.~1b to that of Fig.~1c, 
 we would have to change the asymptotic behavior
of the superpotential $W(x)$ from being even  to being odd at infinity. 
This change of asymptotic behavior  causes vacua to ``appear" (or ``disappear")
from infinity (i.e. the second minimum of Fig.~1c). We will see examples \cite{IT1}
of such behavior when we consider field theory models; see Section 5.2.2.

Finally, we add some comments on the Witten index, Tr $(-1)^F$, in field 
theory. Witten \cite{wittenindex}, \cite{wittennewgauge} 
calculated the index in pure supersymmetric Yang-Mills theory
(i.e. without matter), and found it nonvanishing. 
Thus, pure SYM theory does not break supersymmetry. 
A corollary from Witten's 
result is that vectorlike gauge theories without classical flat directions (or, which
is the same, with added mass terms for all matter fields) also do
not break supersymmetry.
This is because at low energies the  vectorlike theories with massive matter
 flow to pure SYM, for which the index
calculation gives a nonzero result.
This argument can fail in vectorlike theories with classical
flat directions (since 
such theories have a moduli space which renders the index
ill-defined at the classical level) \cite{wittenindex, IT1}. 
Even so, we will see
that most known theories exhibiting dynamical supersymmetry breaking are
chiral.

\section{Supersymmetry breaking in field theory.}

\subsection{The order parameter.}

We start this section by first reviewing some general features associated 
with supersymmetry breaking in quantum field theory.  
The order parameter  for  (global) supersymmetry breaking  is simply the 
vacuum energy. To see this we note that  the
$N=1$ four-dimensional supersymmetry algebra 
\cite{WB}
\beq
 \label{n1susy}
 \left\{ Q_\alpha, ~\bar{Q}_{\dot{\alpha}} \right\} ~=~
 -~ 2 ~ \sigma^\mu_{\alpha \dot{\alpha} } ~ P_\mu ~
\eeq
reduces, in the rest frame of the system, $P_0 = H, \vec{P} = 0$,
after appropriate
rescaling, to the nonrelativistic algebra (\ref{qmsusyalgebra}). Thus, the
arguments following eq.~(\ref{qmsusyalgebra}) can be repeated in the
field theory case, showing that 
the order parameter for supersymmetry breaking is the vacuum energy.

\subsection{Goldstone fermions.}

A straightforward generalization of Goldstone's theorem implies, very generally,
that  if  global supersymmetry is broken there  must be  a  massless fermion 
  in the spectrum,
coupling  to the supercurrent. It is called a goldstino.  
The basic idea is to consider a Green's function, 
$G^{\mu}_{\beta \dot{\alpha}}(x) = 
\langle 0 \vert T S^{\mu}_{\beta} (x)  \bar{\psi}_{\dot \alpha} (0)
             \vert 0 \rangle$, 
  involving the supercurrent, $S^{\mu}_{\beta}$, and a fermionic field 
$\bar{\psi}_{\dot \alpha}$.
Since the current is conserved, we have:
\beq
\label{gthrm}
\int~d^4x ~\partial_{\mu}G^{\mu}_{\beta {\dot \alpha}}~=~ \langle 0 \vert 
~ \{ Q_{\beta}, ~ \bar{\psi}_{\dot \alpha} (0) \} ~\vert 0 \rangle,
 \eeq
(the supercharge is, as usual, the integral of the zeroth component of the
supercurrent, $Q_\beta = \int d^3 x S^0_\beta (x)$), 
or, equivalently in momentum space:
\beq
\label{momgth}
i~P_{\mu}~G^{\mu}_{\beta {\dot \alpha}}(P)\bigg\vert_{P_\mu \rightarrow 0}~=~
 \langle 0 \vert~ 
\{ Q_{\beta},~\bar{\psi}_{\dot \alpha} (0) \} ~\vert 0 \rangle.
\eeq  
The anticommutator above can be non-zero only if supersymmetry is broken.
Further, if it is nonvanishing, we find from eq.~(\ref{momgth}) 
that there must be 
a massless particle in the spectrum,
giving rise to a pole in the Green's
function $G^{\mu}_{\beta {\dot \alpha}}$ at zero momentum. 
By inserting a complete set of states on the right hand side 
 it becomes clear that this particle must be a fermion, 
$\bar{\eta}$,  with 
coupling to the supercurrent 
$\langle 0 \vert S^\mu_\alpha 
\vert \bar{\eta}_{\dot \alpha} \rangle =
 f \sigma^\mu_{\alpha {\dot \alpha}}$.\footnote{Taking the supercurrent
$\bar{S}^\mu_{\dot \alpha}$ instead of $\bar{\psi}_{\dot \alpha}$
 in (\ref{momgth}) and 
using the supersymmetry algebra  
$ \{ Q_\alpha, \bar{S}^\nu_{\dot \alpha} (x) \}= 
- 2 \sigma^\mu_{\alpha {\dot \alpha}}  T_\mu^\nu (x)$, see eq.~(\ref{n1susy}), 
one   can also  relate the vacuum energy density, $E_0$, to 
the goldstino  coupling, $E_0 = f^2$ \cite{wittendsb}.}

There are two kinds of 
multiplets in an $N=1$ theory in $4$ dimensions---chiral 
multiplets and vector multiplets. Let us denote the corresponding fermions
by $\psi$ and $\lambda$. Under supersymmetry these transform as \cite{WB}:
\beq
\label{cht}
\delta_{\zeta} \psi ~=~ i \sqrt{2} \sigma^{\mu} {\bar \zeta} \partial_{\mu} \phi ~+~
                     \sqrt{2} \zeta F~,
\eeq
\beq
\label{vect}
\delta_{\zeta} \lambda ~=~\sigma^{\mu \nu} \zeta F_{\mu \nu}~+~i \zeta D,
\eeq  
where $F$ and $D$ are the auxiliary fields of the chiral and vector multiplet,
respectively, $\phi$ is the scalar component of the chiral multiplet, and $F_{\mu
\nu}$ is the  field strength of the gauge field of the  vector multiplet.
Either of these two kinds of fermions can be present in the  anticommutator 
in (\ref{momgth}). 
The first two terms in eqs.~(\ref{cht}) and (\ref{vect}) cannot 
acquire vacuum expectation values, 
since Lorentz invariance is unbroken. 
Thus the condition for supersymmetry breaking is that some auxiliary
component, either $F$ or $D$, must acquire a vacuum expectation value. 
In general both $F$ and $D$ terms could get such vevs, correspondingly 
the goldstino will generally 
be a combination of the fermions $\psi$ and $\lambda$. 
  
One distinction between bosonic symmetries
and supersymmetry is worth pointing out. 
For a  broken bosonic symmetry,
the   Goldstone boson is associated
with  long wavelength  fluctuations (``spin waves")  along the flat direction  of
the potential  associated  with the global symmetry.  
In contrast, for broken 
supersymmetry,  a goldstino arises even  when the  vacuum is unique.   

\subsection{Simple examples of $F$- and $D$-type supersymmetry
breaking.}

It is useful to begin the study of supersymmetry breaking  
in field theory by studying  a simple
example, called an O'Raifeartaigh model,  which does not involve 
any gauge
fields---in this case supersymmetry breaking will occur because an auxiliary 
$F$ component acquires a vev.  
As we will see below, the low-energy dynamics in more complicated
situations will often reduce to  a model of this type.  
The example we consider here \cite{Nilles},  
has three fields, $\phi_1,\phi_2,\phi_3$,   with a conventional  K\" ahler potential, 
$K = \sum_{i=1}^3 \phi_i^{\dagger}  \phi_i$, and a
superpotential  given by: 
\beq
\label{orsup}
W~= ~m~ \phi_1~ \phi_2 ~+~\lambda~ (\phi_1^2 - a^2)~\phi_3 .
\eeq
The corresponding scalar potential can then be shown to be, using
(\ref{potential}):
\beq
\label{scpot}
V~=~ {\vert m~ \phi_1 \vert}^2~+~ {\vert  \lambda~(\phi_1^2-a^2) \vert }^2
       ~+~{\vert m~ \phi_2 + 2~\lambda~ \phi_3~ \phi_1 \vert }^2,
\eeq
where the three terms on the right hand side are the squares of the 
$F$ components 
of $\phi_2$, $\phi_3$, and $\phi_1$, respectively. 
One can see immediately that  the first two terms on the right hand side
cannot both be zero, thus the vacuum energy must be non-zero and
supersymmetry  is broken.  
If $\vert m \vert^2 > 2 \vert \lambda^2 a^2 \vert$,
one finds that the global minimum lies at  $\phi_1=\phi_2=0$, correspondingly
$F_3$---the auxiliary component of $\phi_3$---acquires a vev.   The potential,
 eq. (\ref{scpot}) 
has  a flat direction  which corresponds to varying $\phi_3$; it can therefore take
any value.   Vacua 
corresponding to different values of $\phi_3$ are physically
different; for example, 
the spectrum  of the theory depends on $\langle \phi_3 \rangle $. 

Since supersymmetry is broken, we do not expect, in general, bosons and their
fermionic partners  to  be equal in mass.  It is easy to see that this 
is in fact true. For example, working in the vacuum where  
$\langle \phi_3 \rangle=0$ one finds the 
 following  excitations. In the bosonic spectrum,  $\phi_3$ has zero mass, 
$\phi_2$ has mass $m$,  and 
two real scalar fields which arise as combinations of $\phi_1$ and 
$\phi_1^{\dagger}$ have masses, 
$\vert m \vert^2 \pm 2\vert \lambda^2 a^2 \vert$. In the fermionic spectrum,
$\psi_3$ (the fermionic partner of $\phi_3$) is massless, while $\psi_1$ 
and $\psi_2$ pair together with a Dirac mass $m$. We see thus that 
the degeneracy between $\phi_1$ and $\psi_1$ is lifted. 
Notice further that there is one massless fermion, $\psi_3$,  it is the 
goldstino. This is accord with the fact that $F_3$ acquired a vev in this 
vacuum.

We saw above that the potential, eq.~(\ref{scpot}), does not uniquely
determine    $\langle \phi_3 \rangle $ and  the   classical theory has  a flat 
direction. Since supersymmetry is broken we expect quatum effects to lift
this flat direction and  to pick out a unique value of 
$\langle \phi_3 \rangle $.  The quatum effects enter through the 
K\"ahler potential which is  perturbatively renormalized.    
The classical vacuum energy (\ref{scpot}) of the O'Raifeartaigh
 model, in the vacuum with $\phi_1 = \phi_2 = 0$ and $\phi_3$---arbitrary,
is:
\beq
\label{vacuumoraiferty}
V_{class}~=~\lambda^2 ~a^4 ~.
\eeq
The leading dependence of the  quantum effects is incorporated  in  
eq. (\ref{vacuumoraiferty})  by  noting that  $\lambda$ is a running 
coupling  that depends on the scale of the  expectation value. 
Since the Yukawa coupling is not asymptotically
free, it increases logarithmically  upon increasing $\phi_3$.
Thus  it  turns out, after a one-loop effective potential calculation is
performed \cite{Huq}, that  the  minimum of the potential is attained
when  $\phi_3 = 0$. 
We note (and we will point out examples later) 
that the stabilization of classical flat directions by  
perturbative corrections to the K\" ahler potential has
important model building applications \cite{Shirman, H2, DDGR, AHM1, GR}.

Finally, we  give an example of $D$-type 
(``Fayet-Iliopoulos type") supersymmetry breaking. $D$-type breaking can
only occur in Abelian gauge theories---it is 
possible to show that (at tree level) supersymmetry
breaking in non-Abelian theories is controlled 
by  F-terms only \cite{WB}.
We will not give other examples of $D$-type breaking in this review. 
We would only like to stress
 that Fayet-Iliopoulos-type supersymmetry breaking may be of
phenomenological relevance. The occurrence of $U(1)$ factors of
the gauge group with Fayet-Iliopoulos terms is common in string theory 
compactifications \cite{DSW}. 
The relevant $U(1)$ factors are usually anomalous (the 
anomalies are canceled by the Green-Schwarz mechanism) 
and generate  Fayet-Iliopoulos 
terms at one loop. The model discussed below illustrates this 
rather  generic mechanism of supersymmetry breaking.

As an example of 
$D$-type breaking we  consider a $U(1)$ supersymmetric
gauge theory with two ``electrons"---two chiral superfields, $Q$ and $\bar{Q}$,
with $U(1)$ charge $+1$ and $-1$, respectively. The K\" ahler potential and 
the superpotential are:
\beqa
\label{fimodel}
K ~&=&~ Q^\dagger~e^V ~ Q ~+~ \bar{Q}^\dagger~e^{- V} ~ \bar{Q} 
~+~2~ \xi_{FI}~V~, ~\nonumber \\
W~&=&~ m~ Q  ~ \bar{Q} ~,
\eeqa
where $V$ denotes the $U(1)$ vector superfield and $\xi_{FI}$ is the 
Fayet-Iliopoulos term (which has dimension of 
mass squared and can be easily seen to be gauge invariant,
see \cite{WB}). The scalar potential of the model is:
\beq
\label{fiscalar}
V~=~ \vert m ~Q \vert^2 ~+~  \vert m ~\bar{Q} \vert^2~ +~ {1\over 8} ~
\vert  Q^\dagger~ Q -  \bar{Q}^\dagger ~\bar{Q} + 2  \xi_{FI}  \vert^2 ~.
\eeq
The first two terms in $V$ are the $F$ terms of $\bar{Q}$ and $Q$, respectively,
while the last term is the square of the $D$ term of the vector multiplet.
It is easy to see from eq.~(\ref{fiscalar}), that if the Fayet-Iliopoulos term
vanishes, $\xi_{FI} = 0$ (and $m \ne 0$) the vacuum occurs for $Q = \bar{Q} = 0$
and supersymmetry is unbroken. 
On the other hand, if both the mass and
the Fayet-Iliopoulos term are nonvanishing, supersymmetry is clearly broken.
The breaking of supersymmetry is  $D$ type if $m^2 > \xi_{FI}/2$ and 
the $U(1)$ gauge 
symmetry is unbroken (the goldstino field then is the gaugino, as is clear from
its supersymmetry 
transformation law, eq.~(\ref{vect})). On the other hand, when 
  $m^2 < \xi_{FI}/2$ supersymmetry breaking is of mixed $F-D$ type
and the gauge symmetry is broken (the goldstino field is then a 
mixture of the gaugino and the fermionic components of $Q, \bar{Q}$  \cite{WB}).
We note that the model discussed above is an example of a model with vanishing
Witten index---it breaks supersymmetry for nonvanishing $\xi \ne 0$, while
for $\xi =0$ supersymmetry is unbroken.

\subsection{Broken global symmetries and supersymmetry breaking.}

It is also useful to comment at this stage on the relation
between $R$ symmetries and supersymmetry breaking. Symmetries which do 
not commute with the supersymmetry generators are called $R$ 
symmetries.\footnote{We use the convention of \cite{WB},  
where the $R$ charge of the superspace coordinates,
$\theta_\alpha$, equals 1. Thus, the fermion component of
 a chiral superfield of $R$ charge $q$ has $R$ charge $q - 1$, the superpotential
in an $R$ symmetric theory has $R$ charge 2, the fermions in vector 
multiplets (gauginos) have $R$ charge 1, etc.} 
Consider a situation where the dynamics responsible for supersymmetry 
breaking can be described by an effective theory involving only chiral 
superfields. We denote these fields by $\phi_i, i=1, ... n$, in the 
discussion below. Supersymmetry is unbroken if 
\beq
\label{cond}
F_i ~=~ {\partial W \over \partial \phi_i} ~=~0~,
\eeq 
for all fields $\phi_i$. Eq.~(\ref{cond}) imposes $n$ holomorphic conditions on 
$n$ complex variables (the $\phi_i$). 
If the superpotential $W(\phi_i)$ is generic, 
it should be possible to satisfy all these conditions and supersymmetry is 
not broken.  

We now investigate how things change 
if the superpotential preserves a global
 symmetry.  For a non-$R$ symmetry one 
can show that that the above argument 
goes through essentially unchanged. 
The global symmetry means 
that the superpotential only depends on appropriate combinations 
of the $\phi_i$ which 
are singlets of the global symmetry.  In terms of these reduced number 
of degrees of freedom, eq.~(\ref{cond}) imposes  an equally reduced number of 
conditions, again leading to unbroken supersymmetry. 

However, 
for an $R$ symmetry
things can be different. 
In this case the superpotential is not invariant 
and  has an $R$ charge $2$.  If the field $\phi_1$ is charged under the $R$ 
symmetry, $W$ can be written as (we are assuming here that 
$\phi_1$ has a expectation value, breaking thus the $R$ symmetry):  
\beq
\label{redef}
W~=~\phi_1^{2/q_1}~f ( X_i ) , ~X_i~=~\phi_i ~ \phi_1^{-{q_i\over q_1}}~.
\eeq
Now, for supersymmetry to be unbroken we have the conditions $W_i = 0$, or,
equivalently:
\beq
\label{co1}
{\partial f \over \partial X_i}~=~0,
\eeq
and
\beq
\label{co2}
f(X_i)~=~0.
\eeq
Notice, that these are $n$ equations but in $n-1$ variables. Generically they
will not be met and supersymmetry is broken \cite{NS}. 

The above discussion leads us to believe that a broken $R$ symmetry is 
necessary for supersymmetry breaking. One way in which this 
conclusion can be avoided is if, unlike what was assumed above, the 
superpotential is  not generic. The superpotential is, 
after all, protected from corrections in perturbation theory 
by a non-renormalization theorem. 
 Corrections can be generated non-perturbatively but these are 
often of a very special form. Thus in several instances the superpotential is 
non-generic and supersymmetry is broken even in the absence of an 
$R$ symmetry. 
Another way in which this conclusion is avoided is if the underlying 
theory does not possess an $R$ symmetry, but the $R$ symmetry arises
as an accidental symmetry in the 
superpotential of the effective 
theory---involving the fields $\phi_i$---discussed above \cite{ISS}, \cite{LRR}.  
Once again, this can happen because non-perturbative effects
lead to  corrections to the superpotential of restricted form.
 In this case the 
$R$ symmetry will be broken by higher dimensional operators
in  the K\" ahler potential.
 In this context we should also mention that upon coupling to  supergravity, 
the continuous $R$ symmetry
is always broken by the constant term 
in the superpotential needed to cancel the cosmological
constant \cite{BPR}.
Finally, our discussion assumed that the relevant effective theory only 
contained  chiral superfields. This is not true in general, as we will 
see below in our discussion of non-calculable models. 
The argument above does 
not apply to such situations, although in several cases of this type 
an $R$ symmetry is present and 
in fact the corresponding 't Hooft anomalies play an important role 
in establishing the breaking of supersymmetry \cite{ADS4}.  

Another relation between broken global symmetries and 
broken supersymmetry is the following \cite{ADS1}:
if the theory has no classical 
flat directions and has a broken global symmetry,
then supersymmetry is broken. To see this,
note that if a global symmetry is broken while supersymmetry is unbroken, the
Goldstone boson of the broken symmetry has a massless scalar 
supersymmetric partner. 
 Since the Goldstone boson is the phase of the order parameter, its
supersymmetry partner corresponds to a dilatation of the
order parameter, and thus 
represents a flat direction of the theory.\footnote{Unless
the low energy theory is described by a compact 
supersymmetric nonlinear sigma model. 
Such sigma models,
however, can only be coupled to gravity if   Newton's 
constant is quantized \cite{BW}. 
Renormalizable gauge theories can be coupled to gravity for all, even
 arbitrarily small, values 
of the Newton
constant. This should  also hold for  
their low-energy effective theories. Thus, we
conclude that the low energy theory of a renormalizable theory 
can not be a  compact supersymmetric sigma model.}
But the theory has no classical flat directions, 
and it is unlikely that strong coupling dynamics 
will lead to their appearance. One thus concludes 
that supersymmetry is broken  in a theory with no classical flat directions
and a broken global symmetry.

\section {Models of dynamical supersymmetry breaking.}

In the previous section, we studied some general features of 
supersymmetry breaking in field theory and  gave some 
  examples, where supersymmetry breaking occurred at
tree level.  As was discussed in the Introduction,  both from the theoretical 
and phenomenological points of view  it is much more interesting to explore the
 problem  of non-perturbative supersymmetry breaking.  
This is the  question  to which we now turn. 

We saw  in our discussion of supersymmetric QCD that  supersymmetric 
 gauge theories  generically have flat directions at the classical level.
Non-renormalization theorems tell us that these directions are not lifted 
in perturbation theory but, as we saw in Section 2, they can be lifted  by 
non-perturbative effects.   The basic idea in dynamical supersymmetry
breaking  is to involve these non-perturbative effects in an essential  way 
in the lifting of flat directions,  leading to a non-zero vacuum energy and thus
supersymmetry breaking. 

There has been a great deal of  research  in  dynamical supersymmetry
breaking in the recent past
and many new 
examples  exhibiting this phenomenon have been constructed 
\cite{ISS}-\cite{LT97}. 
It would be inappropriate to   discuss all of them here.  Instead,
we  will talk about a 
few illustrative examples and  content ourselves by providing  references  to the
rest of the literature.  
In organizing  the  discussion,
 it is useful to begin by  thinking  about the
various energy  scales involved in the problem.  The non-perturbative effects are
characterized by the strong coupling scale, $\Lambda$,  of  a  gauge theory.  
In studying supersymmetry breaking, it is  helpful if the 
scale of supersymmetry breaking can be  made much lower than the
scale  $\Lambda$.  
There are a few good reasons for this.   First,  in many cases, once the two 
scales are
separated, at energies lower than $\Lambda$   the gauge degrees  of freedom  
can  be
integrated out  giving rise to a  much simpler non-linear
sigma model.  
Second,  as was mentioned in Section 2, most of the recent 
advances  in the study of non-perturbative supersymmetric theories have
been restricted to the infra-red, i.e. energies much lower than $\Lambda$. 
Once the   supersymmetry breaking scale  lies in this region, these  powerful 
tools can be brought into play. In this review we will mainly discuss examples
where such a separation of scales can be arranged.   

The separation between the strong coupling scale $\Lambda$ and the
supersymmetry breaking scale  can 
be arranged as follows.  It turns out that in
 many cases  for  supersymmetry breaking to occur,
  a tree level superpotential  is required. 
By  making the coupling constant of these tree level terms  small enough 
the supersymmetry breaking scale can be lowered.  In some cases we
discuss below,  these
terms will be  non-renormalizable and  the corresponding couplings will be 
naturally small (i.e. suppressed by a small ratio of scales). 
 In others, 
we  will  have to adjust some dimensionless Yukawa coupling to be small instead. 

As was mentioned above, once the supersymmetry breaking scale can be lowered,
one can  integrate the gauge degrees of freedom out, at the  scale
$\Lambda$,  giving rise to a non-linear sigma model.  In Section 2,
we  saw that such a sigma model is characterized by  both a K\" ahler
potential and a superpotential.  In all the cases we study, the full 
superpotential  will be determined. However, it will not always  be
possible to determine the K\" ahler potential.   

In Section 5.1, we will first study  ``calculable" models, in which 
the K\" ahler potential can be determined as well.  This will allow us to 
explicitly determine where the supersymmetry breaking vacuum lies 
and ask more detailed questions about it. We study calculable models
where supersymmetry breaking occurs due to instanton-induced
superpotentials (Sections 5.1.1 and 5.1.2) or gaugino condensation
(Sections 5.1.3 and 5.1.4). Finally, we briefly 
mention the calculable models with flat directions (``plateau" models)
in Section 5.1.5.

In Section 5.2, we will then turn to  theories where  the K\" ahler potential
cannot be determined, but where  we will still be able to establish that 
supersymmetry is broken.  In Sections 5.2.1-5.2.3, we consider examples 
that demonstrate how the
techniques of holomorphy and duality come into play in studying supersymmetry
breaking.
Finally, in Section 5.3, we  consider some models where  the scale 
of supersymmetry breaking is  of the same order as  the strong coupling scale.
By using the global symmetries and 't Hooft anomaly cancellation arguments,
we will see how supersymmetry breaking can be established in these cases 
as well. 

\subsection{Calculable models.}

We begin our discussion by considering  models  where  the low-energy 
effective theory responsible for supersymmetry breaking can be completely
determined.  In turn this will allow us  to  explicitly find where the 
supersymmetry breaking vacuum lies,
ask how the other global symmetries of the theory are realized,  and calculate the 
masses of the low-energy excitations.  Because of the detailed information that 
can be extracted from  such calculable models,
  they have played a very useful 
role in  phenomenological  studies, see Section 6. 
They have also served as an important  starting point for constructing
entire new classes of supersymmetry breaking theories. 

We  will consider two examples in detail here. They have  an $SU(3) \times SU(2)$
and $SU(4) \times U(1)$ gauge symmetry, respectively, and are referred to as 
the $(3,2)$ and $(4,1)$ models.  In both cases by adjusting a Yukawa
coupling we will make the supersymmetry breaking scale low  compared 
to the relevant strong coupling scale  of  gauge dynamics.  The K\" ahler 
potential will be calculable in both cases, although the reasons behind this 
will be somewhat different.   We will also comment on various generalizations 
of these examples. 

\subsubsection{Instanton-driven supersymmetry breaking: 
the $(3,2)$ model.}

The $(3,2)$  model was first studied by Affleck, Dine and Seiberg  \cite{ADS1,ADS2}.
It consists of a theory  with an  $SU(3) \times SU(2)$ gauge  group. 
 In addition,  the theory has  two anomaly free  global
symmetries,  $U(1)_Y$ and  $U(1)_R$.   Under these various symmetries  the
matter  content transforms as follows 
($\bar{Q}^i_{~\alpha} ~ \equiv ~(\bar{D}^i,\bar{U}^i )$):\footnote{The  
reader might have noticed that  this theory
is quite similar to the one generation standard model, with  two  differences: 
$U(1)_Y$ is a global symmetry and  the positron field is missing. In fact,
the theory has $U(1)_Y$ and $U(1)_Y^3$ anomalies.}
\begin{equation} \label{Q}
\begin{array}{c|cccc}
&SU(3)& SU(2)& U(1)_Y& U(1)_R \\ \hline
Q^\alpha_i & \Yfund & \Yfund & 1/6&1  \\
\bar{U}^i &\overline{\Yfund} & 1 &- 2/3 &0  \\
\bar{D}^i & \overline{\Yfund} &1& 1/3& 0\\
L^\alpha & 1 & \Yfund &-1/2 & - 3
\end{array}
\end{equation}

Let us first study the classical behavior of this theory. 
The $SU(3)$ and $SU(2)$ D-flatness conditions are given by:
\begin{equation}
\label{flatdirectionsSU3}
Q^{\dagger ~m}_{\alpha}~Q_{l}^{~\alpha}~ - ~\bar{Q}^{m}_{~\alpha}
{}~\bar{Q}^{\dagger ~\alpha}_{l}~ =~ 0,
\end{equation}
and,
\begin{equation}
\label{flatdirectionsSU2}
Q^{\dagger ~i}_{\alpha}~Q_{i}^{~\beta} ~+~
L^{\dagger}_{\alpha}~L^{\beta}~ =~ {1\over2}~\delta^{~\beta}_{\alpha}
{}~( Q^{\dagger}~Q ~+~ L^{\dagger}~L ),
\end{equation}
respectively.  These conditions do not select a unique vacuum,
rather there are  $3$ (complex) flat directions  in the theory which can be
parametrized by the gauge invariant chiral superfields ("moduli"):
\beq
\label{mod32}
X_1~=~Q~{\bar D}~L,~~X_2~=~Q~{\bar U}~L,~~
X_3=~{\rm det}~{\bar Q}_{\alpha}~Q^{\beta}~.
\eeq
At a generic point  along these flat directions the $SU(3)$ and  $SU(2)$
gauge symmetries are completely broken, the corresponding gauge bosons
and their superpartners are heavy, and the low-energy dynamics can be described
by  an effective theory containing  the $X_i$ chiral superfields. 

Now let us turn to the quantum behavior of this theory.  
One finds that  instanton effects lift   the flat directions 
 and give rise to a superpotential in the low-energy effective theory
of the form:
\begin{equation}
\label{wnp32}
 W_{dyn} ~=~{\Lambda_3^7 \over X_3}\; .
\end{equation}
Eq. (\ref{wnp32}) is determined in the following way.  It is the only term that
is allowed by holomorphy and the symmetries of  the theory.   Further, 
 an explicit  (constrained) instanton calculation    shows  that   it does  arise.   
The non-perturbative superpotential, eq.~(\ref{wnp32}),  
gives rise to a potential energy  that
is minimized when some 
fields acquire large expectation values and "run away" to
infinity. Thus we find that the quantum theory does 
not have a stable ground state. 
 
To avoid this problem,  we can  add a tree level superpotential, preserving
the $U(1)_Y \times U(1)_R$ symmetry,  of the
form: 
\beq
\label{wtree32}
W_{tree} ~=~\lambda~ Q \cdot \bar{D} \cdot L ~=~  \lambda~ X_1 ~ .
\eeq
Classically, one now finds that  the $F$ term conditions following from 
this superpotential, along with the $D$ term conditions,
(\ref{flatdirectionsSU3}),  (\ref{flatdirectionsSU2}), lift  all the flat 
directions giving rise to a  unique  vacuum with all the fields set to zero. 
  
The behavior of the quantum theory is much more interesting.   One can show 
that 
in this case  the exact superpotential in the low-energy
effective theory  is  given by a sum of the two terms, eq. (\ref{wnp32}) and
(\ref{wtree32}), to be:  
\begin{equation}
\label{exact32}
W ~ =~ W_{dyn}~+~W_{tree}~=~ {\Lambda_3^7\over X_3} ~+~ \lambda ~X_1~.
\end{equation}
To show this, one uses, following \cite{S1},
 holomorphy, symmetries, and various limits. Note first that  when 
$\lambda \rightarrow 0$   and $X_3 \rightarrow \infty$ the superpotential 
is reliably given by eq.~(\ref{exact32}). 
By holomorphy and symmetries,
the most general form of the superpotential is $W_{dyn}\times f(t)$, with $f$ an
arbitrary function of $t \equiv \lambda X_1
X_3/ \Lambda_3^7$.  Now,  we see that  any value of $t$ can be obtained
by  taking $\lambda \rightarrow 0 $ and $X_3 \rightarrow \infty$
appropriately. Thus,  eq.~(\ref{exact32}) should be  exact.

{}From eq. (\ref{exact32}) it follows that the 
$F$-flatness condition for the field $X_1$,
$dW/dX_1=\lambda=0$ can not be satisfied. Thus one concludes that  
supersymmetry is broken in the quantum theory. 
  
We have assumed  in this analysis
that  an effective theory in  terms of the three fields $X_i$ correctly describes the
low-energy dynamics.  If extra  massless degrees of freedom
 were to enter  the low-energy  theory  at some points in moduli  space,  
the description in terms of the
fields $X_i$  would break down. 
This would manifest itself,  for example,  in 
singularities in the K\" ahler metric $K_{i j^*}$.  
At points where these singularities are present, 
the energy  which goes like $W^*_{i^*} K^{-1 i^* j} W_j$ could go to zero 
and supersymmetry would not be broken. 

In fact, as we will see in a moment, for  small enough $\lambda$
the vacuum  lies in a region of field space where both the $SU(3)$ and 
$SU(2)$ groups are higgsed and weakly coupled. Thus,  
we will be able to explicitly 
compute the spectrum and show that it is consistent with the absence of 
extra massless particles.  But  before doing so,
 it is     worth noting  that
this result also follows quite generally from duality. 
 The $(3,2)$ model (for small enough
$\lambda$) can be shown, by adding extra vectorlike flavors,
 to be dual  to a weakly coupled 
magnetic theory. This dual
theory is completely higgsed and  
one can show that  the  low-energy spectrum 
corresponds to the $X_i$ fields with no additional massless particles.  

Let us now study the supersymmetry breaking vacuum in more detail. 
As was  mentioned at the outset this model is calculable---one can determine
the expectation values of the fields and the low-energy spectrum explicitly. 
We will not be able to provide the full details here, see refs.~\cite{ADS2, BPR}, 
and will content ourselves with  sketching out the  general  picture   and
providing some of  the steps in the calculations. 

The basic idea is  to do a self consistent analysis. One begins by assuming 
that the  field expectation values 
break  both the $SU(3)$ and $SU(2)$ groups  at  energies much above their 
strong coupling scale.  This   allows us to  compute the K\" ahler potential
and  determine the  full non-linear sigma model.  The energy can then be
explicitly minimized to determine these expectation values and 
verify  that the starting assumption was in fact correct.  Before going
further, let us note that the assumption one begins with is  very 
plausible.  
The  non-perturbative
superpotential  pushes the  vacuum out to large field strengths.  
In contrast,
 the tree
level superpotential results in a contribution to the energy that grows at large field
strengths.   The minimum should lie where these two terms  balance each
other (see Fig.~2).  
{}From eq.~(\ref{exact32}) it follows that the  corresponding 
vacuum expectation value, $v$, should roughly go like
\beq
\label{estv}
v ~\sim~ {\Lambda_3 \over \lambda^{1/7}}~.
\eeq
For small enough $\lambda$, $v$ can be made large and if enough fields 
get expectation values both $SU(3)$ and $SU(2)$ should be broken.

\begin{figure}[ht]
\centerline{\psfig{file=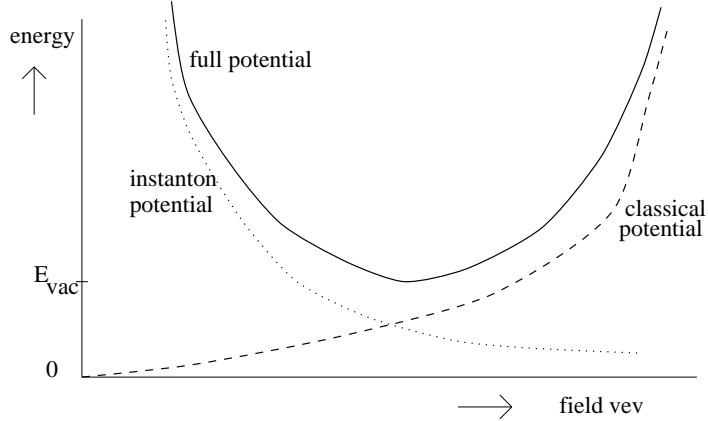}}
\caption{\protect \small The scalar potential in the 3-2 model: instanton vs. tree-level
contribution.}
\end{figure}

The K\" ahler potential in terms of the fields $X_i$ can be calculated. At 
tree level it can be determined to be:
\begin{equation}
\label{kahler32}
K = 24\;{ A ~+ ~B ~x \over x^2}~, 
\end{equation}
where 
\begin{equation}
A ~= ~{1\over 2}\;(\; X^{\dagger}_1 ~X_1 ~+~
X^{\dagger}_2~ X_2\; ), ~~
B ~= ~{1\over 3}~\sqrt{X^{\dagger}_3 ~X_3},~~
x ~=~4~\sqrt{B}\; {\rm cos} \left(~{1\over 3}
\; {\rm Arccos} ~ {~A~ \over B^{3/2}}~\right) \; .
\end{equation}
Although its form is complicated there is a straightforward way to determine 
the K\" ahler potential.  One starts with the canonical K\" ahler potential for the 
$Q, \bar{Q}, L$ fields and  projects onto the D-flat directions,  
(\ref{flatdirectionsSU2}), (\ref{flatdirectionsSU3}),  see~\cite{ADS2}. Equivalently,
one can integrate out  the vector fields that 
become heavy along the flat directions \cite{PR}, \cite{BPR}.\footnote{In the
mathematical literature the above 
 procedure of constructing the tree-level effective theory
of the D-flat moduli is known as the 
``K\" ahler quotient," see, e.g. \cite{HKLR}.} 

Let us stress that  the tree level K\" ahler potential will in general 
receive both perturbative and non-perturbative corrections.  
However,  if  $v$ is large enough, eq.~(\ref{estv}),  and both
the $SU(2)$ and $SU(3)$ gauge groups are broken, these can be neglected.

With the K\" ahler potential and superpotential, eqs.~(\ref{kahler32}), 
(\ref{exact32}),  
at hand,  the non-linear  sigma model is completely determined. 
The energy can now be  found  from eq.~(\ref{potential})
 and minimized. We omit some of the
details here.  On doing so one finds that  the full $SU(3) \times SU(2) $ gauge
symmetry  is indeed broken  at a scale or order $v$, eq.~(\ref{estv}). Thus the
starting assumption is validated.  Furthermore, as expected from the above
discussion,   supersymmetry  is   broken. The vacuum energy is of order $E
\sim \lambda^{10/7} \Lambda^4$, as one expects from eqs.~(\ref{estv}) and
(\ref{exact32}).

It is also worth discussing briefly how the other global symmetries are 
realized in this vacuum. 
It turns out that the $U(1)_Y$ global symmetry
is unbroken in the ground state, while the $R$ symmetry is broken.
The reader might recall that once the tree level superpotential 
(\ref{wtree32}) is added the theory has no flat directions at tree level. 
Thus from the general considerations of Section 4.4 we expect that 
once the $R$ symmetry is broken, supersymmetry is broken as well. 
This is indeed what we have found. 

The massless spectrum consists of a massless goldstino, 
an $R$-axion (goldstone boson
of the spontaneously broken $R$ symmetry), and an additional massless fermion,
with charge $-1$ under $U(1)_Y$ (its existence can be inferred from 
  't Hooft anomaly matching for tr $U(1)_Y$ and tr $U(1)_Y^3$).
Note that all these fields arise from the $X_i$ fields; as promised above,  
there are no additional massless particles. Further,
 all other
components of the $X_i$ chiral superfields have masses of order $\lambda v$.
Finally, by  considering the
full $SU(3)\times SU(2)$ theory, one can also determine the spectrum and
supersymmetry-breaking 
mass splittings of the heavy vector multiplets.

Finally, we note here that our consideration of the $(3,2)$ model
applies to a limited region of parameter space---our considerations are
valid whenever the expectation value (\ref{estv}) 
$v \gg \Lambda_2, \Lambda_3$, i.e. $\lambda \ll 1$ and,   
\beq
\label{validity}
\Lambda_3 ~\gg ~\lambda^{1/7}~ \Lambda_2~.
\eeq
The analysis above showed that the Witten Index vanishes.  This is of course
true more generally, and so the theory could break supersymmetry for other 
values of the parameters as well.  
In fact, for 
$\Lambda_2 \gg \Lambda_3$ (with $\lambda {\rm still \ll 1}$),
 when (\ref{validity}) is not obeyed, 
the description of supersymmetry breaking changes \cite{IT1}, 
but supersymmetry remains broken.
Later we will discuss an example of
 how  sometimes different (in the case described in Section 5.2.4, 
``electric" and ``magnetic") 
descriptions of supersymmetry
breaking are relevant in  different in regions of parameter space. 

\subsubsection{Generalizations of the $(3,2)$ model.}

The $(3,2)$ model has a number of interesting  generalizations.
One can think of  constructing this  model  by starting with  two flavor QCD,
gauging  an  $SU(2)$  flavor symmetry and adding  an extra $L$ lepton
field to cancel anomalies.  Some  generalizations of this construction 
are the $SU(N) \times SU(2)$ models in \cite{DNNS}, the 
$SU(2M+1) \times SP(2M)$ models discussed in \cite{DNNS}  and  in  \cite{IT1},
 the $SU(N) \times SU(N-1)$ models of \cite{PST1}, 
the $SU(N)  \times SU(N-2)$ theories of  \cite{PST2}, and the
models of \cite{LT}. 
While  these are  analogous, in their field content,
to the $(3,2)$ model,  the dynamics leading to 
supersymmetry  breaking  in many of them  is quite different.
We will have more to say about some of them in  the following sections. 

We should also comment on some other  calculable models  in 
the literature, which are analogous to the $(3,2)$ model, and which break
supersymmetry.   One example is the 
 $SU(5)$ model with two
$\Yasymm$ and two $\overline{\Yfund}$ representations. This  $SU(5)$
``two generations" model 
has dynamics  
that is very similar
to that of the 3-2 model \cite{ADS3}. 
Its ground state has been recently analyzed in detail \cite{Veldhuis}. 
Using the recent work 
\cite{DMS} on classifying $N=1$ supersymmetric
gauge theories with a simple gauge group and with $\mu_{matter} < \mu_{adj}$
(where $\mu_{matter (adj)}$ is the index of the matter (adjoint) 
representation), 
one can show that
among the theories with a simple gauge group (including both  classical and
exceptional groups)  and ``purely chiral"
 matter content (i.e. such that no mass terms can be 
 added for any fields), the $SU(5)$ ``two generations" 
model of \cite{ADS3} is the only one with 
 completely calculable 
dynamics.\footnote{We thank W. Skiba for a guided tour of ref.~\cite{DMS} and 
discussions.} 

Another example in this class, which can be constructed using  product groups
 is the $SP(4)\times U(1)$ model of \cite{CSS1}.  Finally,  additional 
  calculable models 
 can be found upon ``deforming" noncalculable chiral 
 models by adding the ``right" amount of 
 massive vectorlike matter (see the discussion in the Section 5.3).

\subsubsection{Supersymmetry breaking by gaugino condensation: 
the $(4,1)$ model.} 

We now turn to considering another example of a calculable model 
with supersymmetry breaking. It is based on a gauge theory with $SU(4) \times 
U(1)$ gauge symmetry \cite{DNNS}, \cite{PT1}. 
The $(3,2)$ model we discussed in the previous section 
was calculable because one could arrange for the vacuum to lie in a region
of  moduli space where the full $SU(3) \times SU(2)$ gauge group
was completely higgsed and weakly coupled. In contrast, as we will see,
in the $(4,1)$ model the $SU(4) \times U(1)$ gauge symmetry  is  only  
partially broken to an $SU(2)$ subgroup  which  gets strongly coupled 
and confines. 
Nevertheless, we will
argue that at low enough energies the resulting sigma model is 
weakly coupled and calculable.

The model has the following matter fields and charge assignments:
\begin{equation} 
\label{Q41}
\begin{array}{c|cc}
&SU(4)& U(1) \\ \hline
A_{\alpha \beta} & \Yasymm & 2 \\
Q_{\alpha} & \Yfund & -3 \\
{\bar Q}^{\alpha} & \overline{\Yfund } &   -1 \\ 
S & 1 & 4 
\end{array}
\end{equation}
In studying the quantum behavior of this theory it is convenient to 
first ignore the effects of the $U(1)$ gauge symmetry. 
The $SU(4)$ flat directions can then be described by the following 
moduli:\footnote{We are 
using a notation where Pf stands for the Pfaffian. For example,
 Pf $A = \epsilon^{ijkl} A_{ij} A_{kl}/8$.}
\beq
\label{mod41}
M ~= \bar{Q} \cdot Q~ \sim~ - 4~, ~{\rm Pf} A ~\sim ~4~,~ {\rm and}~ S ~\sim ~4~,
\eeq
where for later convenience we have also shown the $U(1)$ charges of 
the $SU(4)$ moduli. 
Along a generic flat direction the $SU(4)$ gauge symmetry is broken to 
an $SU(2)$ subgroup. 
There is no matter charged under the unbroken $SU(2)$.
In the quantum theory,
 non-perturbative dynamics in this $SU(2)$ theory leads
to confinement. At scales below the $SU(2)$ confining scale the  confined 
degrees of freedom, e.g., the glueballs and their superpartners,
 can be intergated out. The  effective theory below the $SU(2)$ 
strong coupling scale  involves the 
moduli $M,  {\rm Pf} A$, and $S$. 
Gaugino condensation in the $SU(2)$ theory gives 
rise to a non-perturbative superpotential in this effective theory, proportional
to the scale of the unbroken $SU(2)$:
\begin{equation}
\label{41superpotential}
W~\sim~\Lambda_{SU(2)}^3~=
~ {\Lambda_{SU(4)}^5 \over \sqrt{M~ {\rm Pf} A}}~.
\end{equation}
The $U(1)$ gauge symmetry can be incorporated in the effective theory
by "turning on" its gauge coupling. 
In terms of the moduli, $S$, $M$, and Pf $A$, one finds that the flat directions
of the $U(1)$ D-term potential are described by two moduli, $M {\rm Pf} A$ 
and $S M$. From 
this point onwards the analysis has many similarities with that of the 
$(3,2)$ model. We will consequently only sketch out the details. 

The non-perturbative superpotential, eq.~(\ref{41superpotential}), 
only involves the first modulus, $M {\rm Pf} A$. 
It results  in an energy which is minimized when some 
fields are pushed out to infinity; 
thus the quantum theory has a runaway vacuum. To cure this 
problem we introduce a tree level superpotential:
\beq
\label{wtree41}
W_{tree}~= ~\lambda ~S~ {\bar Q} \cdot Q ~=~ S ~ M~.
\eeq
One can show that this tree level superpotential lifts all 
the $SU(4) \times U(1)$ flat directions. 
The full superpotential in the low energy effective theory  is now
given by the sum of eqs.~(\ref{41superpotential}) and (\ref{wtree41}):
\beq
\label{exact41}
W_{exact}~= ~{\Lambda_{SU(4)}^5 \over \sqrt{M~ {\rm Pf} A}} ~ + 
                ~\lambda ~S ~M.
\eeq
{}From eq.~(\ref{exact41}) it follows that the $F$ term condition for
 $SM$ (which is one of the two moduli) cannot be met and thus supersymmetry is 
broken. 

In fact, the resulting vacuum can be explicitly determined since the 
effective theory is weakly coupled in the relevant region of moduli space.
This might come as a bit of a surprise to the reader.  What about the surviving
$SU(2)$ gauge theory which, as we have mentioned above, has strong dynamics 
associated with it and  confines?  The corrections to the superpotential 
were incorporated in eq.~(\ref{exact41}). In the K\" ahler potential one 
expects non-perturbative effects associated with the strongly coupled $SU(2)$
 to give rise to corrections  that
go  like $\Lambda_{SU(2)} / v$.
Here $v$ is the scale of a typical expectation value, which 
can be estimated by balancing the two terms in eq. (\ref{exact41}) and goes 
like, 
\beq
\label{v42}
v ~\sim ~{\Lambda_{SU(2)} \over \lambda^{1/3}}~.
\eeq 
Thus, for small enough $\lambda$, $\Lambda_{SU(2)}/v \ll 1$ and the 
corrections to the classical  K\" ahler potential are suppressed. 

The resulting ground state and spectrum of low-energy excitations can now  be 
explicitly calculated, in a manner very similar to the $(3,2)$ model. 
We will not go into the details here.  
Let us instead briefly review  the picture of the underlying physics that 
was responsible for supersymmetry breaking. 
One starts with an $SU(4) \times U(1)$ theory in the ultraviolet. 
At a scale of order $v$, eq.~(\ref{v42}), 
this is broken to an $SU(2)$ gauge group. 
The $SU(2)$ theory confines at the scale 
$\Lambda_{SU(2)} \sim  \Lambda_{SU(4)}  \lambda^{2/15}$ giving rise 
to the low-energy sigma model. 
In particular, gaugino condensation in the $SU(2)$ 
theory gives rise to a non-perturbative superpotential in the sigma model. 
Finally, supersymmetry breaking takes place in this sigma model, at a scale, 
$E \sim \lambda^{3/10}~ \Lambda_{SU(4)}$.    

We end with one remark.  The tree level superpotential (\ref{wtree41}), 
which lifts all flat directions, preserves an $R$ symmetry. Thus, provided
the $R$ symmetry is broken, we could have concluded at the outset 
from the general considerations of  section 4.4,
that supersymmetry breaking must occur.  

\subsubsection{Generalizations of the $(4,1)$ model.}

The $(4,1)$ model can be generalized in a straightforward way to an 
entire class of theories with $SU(2l) \times U(1)$ gauge symmetry
and with  matter consisting of a single  antisymmetric tensor representation,
 $A \sim \Yasymm$, $2 l-3$
antifundamentals, $\bar{Q}$, 
one fundamental field $Q$ and $2l-3$ fields, $S_i$,
which are uncharged under the $SU(2l)$ symmetry but carry an $U(1)$ 
charge  \cite{DNNS}, \cite{PT1}. 
In all these theories an $SU(2)$ gauge symmetry is left unbroken. 
Gaugino condensation in this group, together with an appropriate tree level 
superpotential then lift all flat directions giving rise to supersymmetry breaking. 
For reasons analogous to the $(4,1)$ case by adjusting appropriate 
couplings one can arrange for the resulting supersymmetry breaking dynamics
to be governed by a calculable theory. 

In turn, the $SU(2l) \times U(1)$ models can be further generalized. 
The matter content in these theories can be thought of \cite{DNNS}, \cite{PT1}
 as arising by starting 
with an $SU(2l+1)$ theory with an antisymmetric tensor, $A \sim \Yasymm$
and $2l-3$ antifundamentals, $\bar Q$, and breaking the gauge symmetry 
down to $SU(2l) \times U(1)$. 
There  are, of course, other possible breakings of 
$SU(2l+1)$. It is natural to ask if they give rise to supersymmetry 
breaking theories as well. This question was addressed by \cite{LRR}.
They added a heavy adjoint field superfield $\Sigma$ to the theory,
with a superpotential,
$W \sim Tr\Sigma^{k+1}$. This allows the $SU(2l+1)$ to be generically broken to
$SU(2 l+1) \rightarrow U(1)^{k-1} \Pi_{s = 1}^{k} SU(n_s)$,
with $\sum_{s} n_s= 2 l +1$. The matter content 
of the $U(1)^{k-1} \Pi_{s = 1}^{k} SU(n_s)$  theory 
can be obtained by 
decomposing the $\Yasymm$ and $2l-3$ $\overline{\Yfund}$s of $SU(2l+1)$ into
representations of
the unbroken gauge group.  
The authors of \cite{LRR} performed a comprehensive analysis of supersymmetry
breaking in this class of models.  A description of their analysis would take us
far from the objective of this article; we only note that
the deconfinement method of \cite{Berkooz} and
the duality in SQCD with 
adjoint matter and superpotential ${\rm Tr} \Sigma^{k + 1}$ \cite{Kutasov} were
essential in understanding supersymmetry breaking.
Ref.~\cite{LRR} concluded that
for $k=2$, supersymmetry is broken, once appropriate Yukawa couplings
are added to the superpotential (to lift the flat directions), while for $k > 2$,
supersymmetry is generically (for exceptions,
 see \cite{LRR}) not broken.
We see that this construction relates many seemingly different models,
for which supersymmetry breaking can be studied in a unified manner. 
For example, in the simplest case 
for $l = 2, k = 2$, starting with the $SU(5)$ model, one obtains,
choosing $n_1 = 4, n_2 = 1$, the $SU(4)\times U(1)$ model, discussed above,
while choosing $n_1 = 3, n_2 = 2$, the
 $SU(3)\times SU(2)$ model of Section 5.1.1 is obtained.

Many other models of dynamical supersymmetry breaking exhibit 
behavior, similar to the $SU(4)\times U(1)$ model. 
Examples are the $SU(N)\times SU(N-k)$ theories 
\cite{PST1}, \cite{PST2} (see also Section 5.2.3), 
with $k = 1,2$ (whose light spectrum 
has been analyzed in detail along the above lines in \cite{PT2}, \cite{AHMMR}), 
the models of \cite{H1}, and many of the models 
in \cite{DNNS}, \cite{LRR}.

\subsubsection{Calculable models with classical flat directions: 
``plateau" models.} 

These calculable models are
based on some of the models considered in  Section 5.2.2, in 
particular the models based on quantum-modified moduli spaces of
\cite{IT1},\cite{IY}. The models have classical flat directions and
have been shown to break supersymmetry. 
However, the vacuum expectation
value can not be calculated in a controlled approximation.
The models can be made calculable by weakly gauging
a global symmetry and using the perturbative corrections to the K\" ahler
potential to stabilize the flat direction 
\cite{H2}, \cite{DDGR} at a large expectation value, in a 
variant of the ``inverse hierarchy" mechanism \cite{WittenIH}.

\subsection{Noncalculable models.}

The calculable models  are  the simplest class of theories  breaking
supersymmetry.  They form an important starting point in the study of 
this phenomenon, and as we have seen, teach us a lot.  
In  pursuing this study further, we would like to complicate things in stages.  
Accordingly,  in this Section, we  continue  to study theories in which by 
adjusting a parameter,  the 
supersymmetry breaking  scale  is  made lower than  the underlying scale of
non-perturbative dynamics.  Thus below the strong coupling scale, but
above the supersymmetry breaking scale we can  use  a supersymmetric effective
theory to describe the dynamics.    But  here, unlike the previous examples,  we  
will    consider situations in which  the non-perturbative effects are  important in
both correcting the superpotential and the K\" ahler potential.   We see  below how
in many cases one  can still argue that supersymmetry is broken. 
But in the absence of  more  information about 
the K\" ahler potential one cannot  find in detail  where the vacuum lies. 

In Section 5.2.1 we consider a simple model,
 where supersymmetry breaking is due to confinement. 
Section 5.2.2 is devoted to 
models of supersymmetry
breaking with classical flat directions that are lifted by nonperturbative
effects. 
In Section 5.2.3, we give an example of how duality can be used 
to study supersymmetry breaking.

Finally, in Section 5.3, we  turn to  the  case where the scale of 
non-perturbative dynamics  and supersymmetry breaking are comparable. 
This is of course the generic situation. 
The full  complexity  of these theories
 makes a controlled analysis  difficult. 
Even so, we show how  in some cases the various global symmetries
and associated 't  Hooft anomalies, and criteria like the Witten index make it 
quite plausible  that supersymmetry is broken.

\smallskip

\subsubsection{Supersymmetry breaking by confinement: the ISS model.}

The first model we  look at was  studied  by Intriligator, Seiberg and Shenker
 \cite{ISS}. Its simplicity makes it a good point to begin. 

The model is an  $SU(2)$ gauge theory with a single
chiral superfield in the 
three-index symmetric representation (i.e. the four-dimensional 
``spin-3/2" representation).  This theory is chiral---no 
holomorphic 
mass term can be written for the single spin-3/2 representation.   To
see this, 
denote the matter field by $q_{\alpha \beta \gamma}$
(with $q_{\alpha \beta \gamma} = q_{\beta \alpha  \gamma} = \ldots$).
It is easy to see then, that the quadratic invariant $q \cdot q$ (indices contracted
with $\epsilon$ symbols) vanishes identically, because 
of the symmetry of $q$. There is no cubic invariant; the only independent
invariant
is then\footnote{More precisely, $u = 
q_{\alpha_1 \beta_1 \gamma_1} \epsilon^{\alpha_1 \alpha_2} \epsilon^{\beta_1
\beta_2}
q_{\alpha_2 \beta_2 \gamma_2} \epsilon^{\gamma_1 \gamma_3}
q_{\alpha_3 \beta_3 \gamma_3} \epsilon^{\alpha_3 \alpha_4} \epsilon^{\beta_3
\beta_4}
q_{\alpha_4 \beta_4 \gamma_4} \epsilon^{\gamma_2 \gamma_4}$.} $u = q^4$.

The theory thus has  a one dimensional moduli space and  along the
flat direction parametrized by $u$, the $SU(2)$ symmetry is totally broken. 
Classically,  the
moduli space is singular at the origin: the
classical K\" ahler potential is 
 $K \sim q^\dagger q \big\vert_{D-flat} \sim (u^\dagger u)^{1/4}$ and the 
K\" ahler metric is singular at $u = 0$.
When $u \rightarrow 0$ the $SU(2)$ symmetry is restored and extra vector 
multiplets become massless. 

The quantum theory is  asymptotically free,\footnote{For 
the spin-3/2 representation,
$T(R) = 5$,  the one-loop beta function of the gauge coupling
is then $b_0 \sim 3 T(G) - T(R) = 1$.} and
 one expects non-trivial dynamics 
in the infra-red. It is also useful to note  that  the  model
has an anomaly free $R$  symmetry, 
under which the field $q$ has charge $3/5$.

The authors of ref.~\cite{ISS} argued that  quantum-mechanically,  the 
theory confines  in the vicinity of  the origin  of moduli space.  As a result,
the classical  singularity  at the origin  is smoothned out without the
appearance of any new massless particles.   
While one cannot prove this assertion,
it meets one non-trivial check. 
At the origin,  the global $U(1)_R$ symmetry mentioned above  is 
unbroken.  The $u$ field  saturates the  't Hooft anomaly   matching conditions
for this $U(1)_R$ symmetry.  This is easy to see: 
the relevant anomalies are
Tr$R = 7/5$ and Tr$R^3 = (7/5)^3$; these are obviously saturated
by the fermionic component of $u$  which has  $R$ -charge $7/5$. 

In the following discussion we will accept the above assertion  that the 
effective theory in terms of the field $u$ is valid everywhere in moduli 
space. It  follows then that the 
K\"ahler potential  is smoothened out near the origin and can be approximated as
 $K \simeq u^\dagger u |\Lambda|^{-6}$, for $u \ll \Lambda^4$.
We now add a tree-level superpotential to the theory:
\beq
\label{wtreeiss}
W_{tree} ~=~ { u \over M_{UV}}~.
\eeq
 This term is nonrenormalizable;
the theory therefore 
should be considered as a low-energy effective
theory valid at  at scales below  $M_{UV}$. 
We note that  this term lifts 
the  one classical
 flat direction in this theory; it also breaks the $R$ symmetry.\footnote{
As an aside,  we note that  the superpotential in the  effective low energy theory,
eq.~(\ref{wtreeiss}), has an accidental $R$  symmetry, 
which is a combination of the $R$ symmetry of the microscopic theory and
the accidental $U(1)$ of the low-energy theory; this accidental symmetry
is broken by higher dimensional terms in the K\" ahler potential.}
In the presence of eq.~(\ref{wtreeiss}) it  follows that the $F$ term condition
for $u$ cannot be met and supersymmetry is broken. 

A few comments are in order at this stage.   First,  we cannot say with 
certainty where the vacuum lies.  In the vicinity of the origin the leading term in 
the K\"ahler potential  $K\simeq u^{\dagger} u |\Lambda|^{-6}$ gives rise to 
an energy that goes like 
\beq
\label{energyiss}
E_0 ~= ~K^{-1}_{u^* u}~ | W_u |^2 ~\simeq ~
{| \Lambda |^6 \over |M_{UV}|^4}~, 
\eeq
and  behaves like a constant.  Thus,  
in determining the  minimum, higher terms in the
K\" ahler potential, which are difficult to estimate,  need to be kept. 
Second,  regardless of exactly where  the minimum lies, 
 eq.~(\ref{energyiss})  gives a good estimate of the energy.   
So, the 
supersymmetry breaking scale is smaller than $\Lambda$,  justifying the 
use of the effective theory  in terms of  $u$.
Finally,   as  was mentioned   above,  the
microscopic theory  one starts with breaks down  at the scale $M_{UV}$.   
However,
by  taking  $\Lambda \ll M_{UV}$, one can be  quite  sure that there is  a
supersymmetry breaking  minimum 
in the region where the effective theory is
valid.  
This is because classically the superpotential, eq.~(\ref{wtreeiss}), lifts all the
flat directions. It follows  then that in the region 
$\Lambda  \ll  (u)^{1/4} \ll 
M_{UV}$,  where  the classical approximation is trustworthy, the energy must 
rise,
leading to  the conclusion that a local minimum must lie in the region 
$(u)^{1/4} \le O(\Lambda)$. 
 It is, of course, possible that the full 
theory has a global supersymmetry 
preserving minimum with $(u )^{1/4}\ge M_{UV}$, but that can not be decided 
without knowledge of the underlying theory at scales above $M_{UV}$.  

The mechanism by which this model breaks supersymmetry is, in fact, more
general and occurs in more complicated theories. For example, the 
$SU(7)$ ''s-confining" theory with matter consisting of two 
sets  of $\Yasymm + 3 \times \overline{\Yfund}$ breaks 
supersymmetry after addition
of an appropriate tree level superpotential
 \cite{CSS2}.

\subsubsection{Models with classical flat directions: the ITIY model.}

We now turn to discussing the theory first studied by Intriligator and
Thomas \cite{IT1}, and Izawa and Yanagida \cite{IY}.
  This model has two remarkable features. 
First, it breaks supersymmetry even though it is non-chiral. Second,
the classical theory has flat directions, which are lifted 
by non-perturbative  quantum effects leading to supersymmetry breaking. 

The ITIY model is an $SU(2)$ gauge theory
with $4$ fundamentals $Q^i$ and six singlets $S_{ij}$, with a tree-level
superpotential,
\beq
\label{spiyit}
W~= ~\lambda ~S_{ij}~ Q^{i} \cdot Q^{j}~. 
\eeq
For further reference we  note that the global symmetries of the theory include 
an $SU(4)$ flavor symmetry under which the $Q^i$ transform as a $\Yfund$ 
and the $S_{ij}$ as an $\overline{\Yasymm}$.
 
Let us begin by studying the classical behavior of this theory. 
The  $SU(2)$ D-flat directions can be described
by gauge invariant chiral superfields, ``mesons," which we denote by 
$M^{ij}  \sim Q^{i} \cdot Q^{j} $.
There are six of these meson fields but they are not all independent;
 classically they satisfy  a constraint:
\beq
\label{clcons}
\epsilon_{ijkl}~ M^{ij}~ M^{jk} ~=~0~.
\eeq
It follows from the superpotential, eq.~(\ref{spiyit}), that 
all the meson  flat directions are lifted,  since the $F$-term 
equations for the $S_{ij}$ fields set all the mesons to zero. 
However the singlet flat directions
remain unlifted. Along these flat directions, 
the mesons are zero but the singlet
fields $S_{ij}$ are free to vary. 

Let us now turn to the quantum behavior of this theory. 
 As in the classical  case, the quantum dynamics is described by the 
mesons $M^{ij}$ and the singlets $S_{ij}$. The difference is that in 
the quantum case non-perturbative effects modify  the constraint 
eq.~(\ref{clcons}) by the addition of a 
term dependent on the strong coupling scale of the 
$SU(2)$ gauge theory,\cite{S1} \footnote{This  was 
discussed in Section  2. In the language of supersymmetric QCD 
we have a situation with  $N_c=N_f=2$  here.} $\Lambda$.
This gives rise to the following full  superpotential 
in the effective theory:
\beq
\label{full}
W_{eff}~= ~\lambda ~S_{ij} ~M^{ij}~ + ~A~\left( \epsilon_{ijkl} ~M^{ij} ~M^{jk}~-~
                                            \Lambda^4 \right) ~, 
\eeq
where $A$ is a Lagrange multiplier that implements the constraint. 

We now find that the theory breaks supersymmetry! The $F$ term conditions for 
the $S$ fields still set all the mesons, $M^{ij}$, to zero,  but now this 
is in conflict with the quantum modified constraint (which follows from 
 the $F$-term condition of the Lagrange multiplier $A$). 

The argument above tells us that supersymmetry is broken, but it 
does not tell us where the resulting ground state lies. We saw above that 
classically the theory had flat directions. One might wonder if 
the energy goes to a minimum (which is non-zero) at infinity, resulting in 
an unstable  runaway ground state. 
In fact, one can show that this does not happen. 
Only the flat directions in the classical theory  are relevant for this 
discussion---along the other directions the energy grows very large as we 
go out to infinity and one does not expect quantum effects to turn this around. 
Consider such direction along which the $SU(4)$ global 
symmetry is broken to $SP(4) \equiv SO(5)$ by giving a vev 
\beq
\label{fltdir}
S_{ij}~=~s ~J_{ij}~,
\eeq
where $J = {\rm diag} \{ i \sigma_2, i \sigma_2 \}$  is the $SP(4)$ invariant tensor. 
If $s$ is large  
all the quarks get a large mass, of order 
$\lambda s \gg \Lambda$, and can be  integrated out at a scale 
much above $\Lambda$.
At low-energies this gives rise
to a theory which contains an $SU(2)$  gauge field (and superpartners) 
with no matter, and the fields $S_{ij}$.  
The strong coupling scale of the low-energy theory is
determined by threshold corrections to be:
\beq
\label{match}
\Lambda_{low}^6~=~\Lambda^4 ~\lambda^2 ~s^2.
\eeq
Gaugino condensation in the $SU(2)$ group now gives rise to  a superpotential,
which goes like $\Lambda^3_{low}$. Substituting from eq.~(\ref{match}) then 
gives:
\beq
\label{wrun}
W~=~\Lambda^2 ~\lambda ~s.
\eeq
The energy along this direction can now be calculated; it is given by:
\beq
\label{seenergy}
E ~=~ \bigg\vert {\partial W \over \partial s} \bigg\vert^2 ~\sim ~
\vert \lambda  \Lambda \vert^2.
\eeq
We see that the energy is a constant, independent of $s$.
 In this discussion, we have so far assumed that the K\" ahler
potential for $s$ is classical. 
In fact this is true, to leading order, but there are 
small corrections \cite{Shirman}, \cite{AHM1}, \cite{DDGR}.
For large enough $s$, 
the leading corrections arise because the coupling $\lambda$ 
is dependent on $s$, and being non-asymptotically 
free increases with $s$ logarithmically. 
Thus from eq.~(\ref{wrun}) we see that the energy 
increases (although only logarithmically) with $s$
 and the runaway behavior is  avoided.\footnote{As $s$ increases, 
at some point one hits a Landau pole  singularity and 
other degrees of freedom must come into play. However this 
scale can be made much higher than $\Lambda$.
 Also, we note that for the 
present purpose the running of the coupling $\lambda$ is governed by the 
$\beta_{\lambda}= 8 \lambda^3/(16 \pi^2)$.} Along other directions
where $S_{ij}$ gets a vev breaking the flavor symmetry to $SU(2) \times SU(2)$ 
one finds similarly that the energy increases and there is no runaway behavior. 

The running of the Yukawa coupling thus 
 pushes the expectation 
value of $s$ to small values, where the theory is not weakly coupled. 
This makes it difficult to reliably calculate the ground state of the model. 
We should mention that it
 is possible \cite{H2}, \cite{DDGR}, 
to stabilize the vacuum at large expectation values upon gauging
(part of) the flavor symmetry of the model, by a variant of the inverse
hierarchy mechanism \cite{WittenIH}. 
This results  in a class of
 calculable models, the ``plateau" models  mentioned in Section 5.1.5.

We began the discussion of the ITIY model by noting that it was 
not chiral.  The reader might wonder how this is consistent with the 
breaking of supersymmetry.
Specifically, one can add mass terms for both the quarks and the $S_{ij}$
fields. This lifts all the  flat directions.  For large enough masses,
the low-energy theory is a pure $SU(2)$ Yang-Mills which has two 
vacua and a Witten index of $2$ \cite{wittenindex}. 
What happens now when the masses are 
taken to zero? On adding mass terms to the superpotential and incorporating
the non-perturbative constraint  we find: 
\beq
\label{wmass}
W~=~\lambda ~S_{ij} ~M^{ij} ~+~ m_{ij} ~M^{ij} ~+~ 
{1 \over 2}~{\tilde m}~ {\rm Pf} S ~+~
A~\left( {\rm Pf} M~ -~ \Lambda^4 \right)~.
\eeq
{}From here it follows that the expectation values are given by:
\beq
\label{vevM}
M^{ij}~ 
\sim ~ \epsilon^{ijkl} ~m_{kl} 
~\left( {\Lambda^4 \over {\rm Pf} m } \right)^{1 \over 2}~,
\eeq
and,
\beq
\label{vevS}
S_{ij}~ \sim ~{m_{ij} \over \tilde m} ~
\left( {\Lambda^4 \over {\rm Pf} m} \right)^{1 \over 2}.
\eeq
The square root above can take two values---this corresponds to the 
two vacua of $SU(2)$ SYM we expect. 
Now we can take ${\tilde m}$ and $m_{ij}$
to go to zero (keeping the relative ratio of masses fixed). We see that 
$M^{ij}$ has a finite limit, but $S_{ij} \rightarrow \infty$. Thus the 
supersymmetry preserving vacua run off to infinity in the limit of 
vanishing mass.  The Witten index changes discontinuously 
from $2$ to $0$, 
because the mass terms change the behavior of the Hamiltonian 
at large field strengths. 

We close this section by noting that the ITIY model has several 
generalizations, see \cite{IT1};
 for simplicity we have focussed on the simplest example of this class here.  

\subsubsection{Supersymmetry breaking and duality: the $(5,3)$ model.}

In this section, we will consider an example of the $SU(N) \times  SU(N-k)$
 models. They 
 were studied in  \cite{PST1,PST2}, where it was shown that 
a large class of these theories 
broke supersymmetry.  As was mentioned in 
Section 5.1.2,  in terms of their matter content,  
these models   can be thought of  as
generalizations  of the  the $(3,2)$ model. 

Here, for  simplicity,  we will discuss a particular model in this class.
It  has  an $SU(5) \times SU(3)$ gauge symmetry.  
In the following discussion we will often refer to this theory as the electric
theory.   We will also construct another  theory, based on  an $SU(5)
\times SU(2)$ gauge group,  which we  call   
the ``magnetic" dual  theory.  
The electric and magnetic theories   
will  be  equivalent to each other in the infra-red
and both  are   useful in learning about  the low-energy  behavior, especially
supersymmetry breaking. 
 In particular, it will be  interesting  to study 
the behavior of the theory  by varying a Yukawa coupling.  
The electric theory  will  yield a calculable  theory of  supersymmetry breaking
in  some restricted  region of parameter space.  
In contrast, the magnetic description
will  not be calculable,  
but it will allow us to establish that supersymmetry is broken
in a much  larger region of parameter space. 

The matter content of  the $SU(5) \times SU(3)$  theory is given as follows: 
\begin{equation} 
\label{53model}
\begin{array}{c|cc}
&SU(5)& SU(3) \\ \hline
Q_{\alpha \dot{\beta}} & \Yfund & \Yfund \\
L^{\alpha i} & \overline{\Yfund} & {\bf 1} \\
R^{\dot{\beta}}_a & {\bf 1} &    \overline{\Yfund } 
\end{array}
\end{equation}
where $i = 1,2,3$ and $a = 1, \ldots 5$ are flavor indices under the global
$SU(3)_L \times SU(5)_R$ symmetry, while dotted (undotted) greek 
letters denote the indices under the $SU(5)$ ($SU(3)$) gauge groups.
 The  theory has a 
renormalizable tree level superpotential:
\beq
\label{wtree53}
W_{tree} ~=~ \tilde{\lambda}_i^a ~ R_a \cdot Q \cdot L^i ~+
 ~ \tilde{\alpha}_{a b} ~ R_c
\cdot R_d \cdot R_e ~\epsilon^{a b c d e} ~.
\eeq
This superpotential lifts all classical flat directions and preserves
a  diagonal, anomaly-free 
$SU(2)$ subgroup of the global flavor symmetry, provided
 the superpotential couplings are of the form:
\beqa
\label{alpha53}
\tilde{\lambda} ~=~\left(  \begin{array}{ccc}
  \lambda & 0  & 0   \\
0 &  \lambda & 0  \\
 0  &  0  & \lambda   \\
 0  &  0  & 0 \\
 0 & 0  & 0 
\end{array}  \right) , ~  ~ ~ ~
\tilde{\alpha} ~=~\left(  \begin{array}{ccccc}
  0 & \alpha  & 0  & 0  & 0 \\
-\alpha &  0 & 0  & 0  & 0 \\
 0  &  0  & 0  & \alpha & 0 \\
 0  &  0  & - \alpha & 0 & 0 \\
 0 & 0  & 0  & 0 & 0
\end{array}  \right)~.
\eeqa
Eq.  (\ref{wtree53}) with the couplings (\ref{alpha53})
can 
be shown to preserve a flavor-dependent, anomaly free  $R$ symmetry.\footnote{
To see this one can start by assigning 
different $R$ charges to $Q$, $L$, $R_{a < 5}$, and $R_5$.
These charges have to satisfy four conditions: two ensuring that the
superpotential terms are invariant, and two---that the symmetry
is anomaly free; it is easy to see then that there is a solution
with nonvanishing charges of all fields.}
Thus, since there are no flat directions, according to the general criterion of
Section 4.4, 
one expects that if the $R$ symmetry is broken 
supersymmetry is also broken.

This is indeed what we will 
 find below when  we  try to understand
 the supersymmetry breaking dynamics in more detail. 
Since the analysis  is  quite  involved,  it is useful to first sketch out  in words
the general idea. 
Throughout this discussion we consider the 
theory  for  $\Lambda_3 \gg \Lambda_5$. We then vary the parameter
$\alpha$ and ask about  supersymmetry breaking.\footnote{ In general,  the 
behavior  of this theory  also  depends on $\lambda$.  For simplicity, we 
will keep  $\lambda$ fixed of order $ \le 1$  here.}
 It  will turn out that 
the magnetic theory  mentioned above will allow us to establish,   for 
$\alpha < 1$,  
that  supersymmetry  is  broken.  However, the magnetic
theory is not calculable and we will not be able to learn much  more 
about the resulting ground state. 
When $\alpha \ll 1$ though,
we will see that the  vacuum lies in a region of moduli space where the electric
theory provides  a weakly coupled sigma model.  
By using this effective theory
we  will   independently  be able to  see that  supersymmetry is
broken and  also learn  a great deal  about the resulting ground state. 

\smallskip

{\flushleft{\it Calculable limit: ``electric" description of supersymmetry breaking.}}

Since the calculable limit is simpler, we start with the situation $\alpha \ll 1$
and first consider the electric theory. 
To understand this case 
it is useful to first consider what happens when $ \alpha=0$. The 
``baryonic term" in eq.~(\ref{wtree53}) 
is absent in this limit and the classical theory has flat 
directions.
 Let us consider one such direction along which the  baryon, 
 $b^{45} = \epsilon^{45abc}R_c \cdot R_d \cdot R_e $  acquires an 
expectation value.  
Along this direction $R^{\dot \alpha}_a=v \delta^{\dot\alpha}_a$, 
$a=1,2,3$, and  the   $SU(3)$ gauge symmetry is completely broken. 
We will  be interested  in what happens  for large values of $v$. 
In particular we will assume that $v  \gg \Lambda_3, \Lambda_5$  and  discuss  
the low-energy effective theory  in this region of moduli space.  
 The supersymmetry breaking vacuum will then lie in this region thereby  
making 
our analysis self-consistent. 
 If $v \gg \Lambda_3$,  the  $SU(3)$ gauge symmetry is broken  while  it is
still weak and  non-perturbative effects  coming from it 
can be neglected.   Furthermore, the  Yukawa coupling---the 
first term in eq.~(\ref{wtree53})---gives  a  large mass,
$\sim \lambda v$   to  the $Q$ and $L$ fields, 
which transform under the  $SU(5)$
gauge symmetry.  Thus,
  these matter fields can be integrated out, leaving a  pure
$SU(5)$ group at low-energies.  
The strong coupling scale of this  theory is given by 
$\Lambda_{5L}^{15} =\lambda^3 b^{45}\Lambda_5^{12}$.   Gaugino
condensation in the low-energy pure $SU(5)$ theory now gives rise to a
superpotential:  
\beq \label{wexact53}
W_{eff} ~\sim ~ \Lambda_{5 L}^3 ~\sim 
~ \lambda^{3 \over 5} ~ \Lambda_5^{12 \over 5} ~ (b^{4 5})^{1 \over 5}~.
\eeq
It is easy to see that this superpotential results in  runaway behavior,
with the $R$ fields being pushed out to infinity.   The behavior described 
above along the $b^{45} \ne 0$ direction is in fact  true along a general 
baryonic flat  direction, $b^{ab} \ne 0$ as well. 

We now 
return to  the original theory, with $ \alpha$, eq.~(\ref{wtree53}),~(\ref{alpha53}),
non-zero but small. 
As was mentioned above,  the  flat directions are now all lifted with the 
choice (\ref{alpha53}).  
However,
 if  $ \alpha$  is small enough we still
expect the  vacuum to lie   far out 
 along the baryonic directions.  The low-energy
effective theory  is then described by  
an independent set  made out of the baryon
fields, $b^{ab}$, 
and  is calculable for reasons analogous to the $(4,1)$  model case studied in 
Section 5.1.3.   
In particular, gaugino condensation in the unbroken $SU(5)$ group 
gives rise to a non-perturbative superpotential  of the form (\ref{wexact53})
in this theory. The 
resulting sigma model  can be  explicitly analyzed,   \cite{PT2}, and shows  that 
supersymmetry is indeed broken.  We will not go into the details here. 

Instead let us  only note that  the expectation values for the $R$ fields can be 
estimated by balancing the non-perturbative term,  eq.~(\ref{wtree53}) with the
second term in the tree level superpotential,
 ${\tilde \alpha}_{ab} R_c \cdot R_d 
\cdot R_e \epsilon^{abcde}$.  This gives an estimate  for $v$:
\beq
\label{vev53}
v ~\sim ~\left( {\lambda^{3} \over \alpha^{5}} \right)^{1\over 12} ~ 
 \Lambda_5~.
\eeq
The supersymmetry breaking scale  then goes like: 
\beq
\label{energy53}
M_{SUSY} \sim ( \lambda^3 \alpha )^{1/12}~ \Lambda_5
\eeq
Finally, the scale  at which the low-energy pure $SU(5)$ theory confines
is  
\beq
\label{lambdal5}
 \Lambda_{5L}  \sim  
\left( \lambda^{3}\over \alpha \right)^{1/12} ~\Lambda_5~.
\eeq
We can now check that the assumptions in the above analysis are consistent. 
For small enough $\alpha$,  $v \gg \Lambda_3, \Lambda_5$; moreover  the 
vacuum energy,  $M_{SUSY} \ll \Lambda_{5L} $. Thus,
the breaking of supersymmetry could be studied in a low-energy  
effective theory  which neglects the non-perturbative effects of the 
$SU(3)$ group and incorporates them  for the $SU(5)$ group as discussed
above.

Now  let  us ask  what happens when  $\alpha$ is increased. We see from 
eq.~(\ref{vev53}) 
that as $\alpha$ increases, $v$ decreases.  Thus at some point,
while $\alpha$ is still much less than one,  we  come to a situation where 
$\Lambda_5 \ll v \sim  \Lambda_3$.   
At this stage we can no longer 
reliably use the  description  above. In particular,
 we cannot neglect the 
dynamical effects in the $SU(3)$ gauge theory.  

\smallskip

{\flushleft{\it The ``magnetic" description of supersymmetry breaking.}}

We now turn to constructing the  dual theory with  $SU(5) \times
SU(2)$ gauge symmetry.   This description will  not be calculable,
but it allows   one to show
quite generally, as long as $\alpha <1$, that supersymmetry is broken.

Before  discussing the dual theory it is important to state  one assumption. 
Strictly speaking,    so far,
Seiberg's duality  has  only been been used  to  relate the electric and magnetic
theories at zero momentum. 
Here we will  assume that the two are equivalent for 
some  range  of  non-zero momentum as well.   
This is not unreasonable---the two
theories should approximate 
each  other for small enough values of momentum.  As
long as this is true our analysis  will be  valid---the 
supersymmetry  breaking can  be brought within this range by tuning the
parameters $\lambda, \alpha$. 

Note that, as was described  at the beginning,
we take   $\Lambda_3 \gg \Lambda_5$. Thus,
  it is useful  in   constructing
the dual theory to first  ``turn off" the $SU(5)$ coupling. 
 In addition, to begin,
we disregard the  tree level Yukawa couplings, eq.~(\ref{wtree53}). 
 Since the $SU(3)$ theory
has now $5$ flavors, i.e. $N_f = N_c + 2$, 
the appropriate description of the infrared physics
 is in terms of a dual 
$SU(2)$ theory, with the following matter content:
\beq
\label{dual53}
\begin{array}{c|cc}
&SU(5)& SU(2) \\ \hline
q^{\alpha}_{\dot{\beta}} &\overline{ \Yfund} & \Yfund \\
r^{\dot{\beta} a }& {\bf 1} &    \overline{\Yfund }\\
M_{\alpha a} & \Yfund & {\bf 1} \\ 
L^{\alpha i} & \overline{\Yfund} & {\bf 1} \\
\end{array} ~ 
\eeq
and a  superpotential
\beq
\label{wdual53}
W ~= ~ M_{\alpha a} ~r^a \cdot q^\alpha ~.
\eeq
Now we turn back on the ``spectator" $SU(5)$ coupling and
 observe that in this dual 
description, the $SU(5)$ theory has five flavors of quarks ($M_a$)
and antiquarks ($L^i, q^{\dot{\alpha}}$).
 It is therefore confining,  with a 
quantum modified moduli space (supersymmetric QCD with $N_f = N_c$).
Below the confining scale of the $SU(5)$ theory, the appropriate 
degrees of freedom are the baryons and mesons: 
\beq
\label{dual53Adof}
N_{a {\dot{\alpha}}}~ \sim ~M_a \cdot q_{\dot{\alpha}} , ~
 K_a^i ~\sim~ M_a \cdot L^i, ~ B ~\sim~ {\rm det} M ,
~ \bar{B} ~\sim ~q^2 \cdot L^3 ~.
\eeq
Hereafter we omit various  scale factors that appear in the duality map; for
details, see \cite{PST2}. The superpotential (\ref{wdual53}) of the theory 
then becomes:
\beq
\label{wdual53B}
W ~ =~ N \cdot r ~+ ~ A ~\left( N^2 \cdot K^3 - B \bar{B} - 
\bar{\Lambda}_{5}^{10} \right) ~,
\eeq
where $A$ is a Lagrange multiplier enforcing the quantum modified constraint. 
The scale $\bar{\Lambda}_{5}$ is the scale of the $SU(5)$ theory
in the dual; it can be found using the duality scale matching and
the symmetries of the problem. 
Below that scale, the appropriate degrees of freedom are 
the mesons and baryons (\ref{dual53Adof}) and the $SU(5)$ singlets 
$r_a$.
We see that now, upon crossing the $\Lambda_{5L}$ threshold, 
the matter content of the $SU(2)$ theory has changed: 
 the $SU(2)$ theory has now $5$ flavors, $N_a$ and $r^a$. 
These flavors, however, are massive: below the scale $\bar{\Lambda}_{5}$, 
 the Yukawa coupling in (\ref{wdual53}) turns into a mass term.
Thus, at low enough energies, the $SU(2)$ theory confines as well.  

The following analysis to find the confined degrees of freedom
in this low-energy theory is straightforward, but rather tedious,
 and we give only the main points, omitting
various details. For simplicity, let us
collectively denote the mesons of the $SU(2)$ theory by $V_{ij}$, with 
$i, j = 1,...,10$
(thus, the matrix $V$ is antisymmetric and has elements 
$N_a \cdot N_b, N_a \cdot r^b, r^c \cdot r^d$).
Along the flat directions of the $SU(2)$ theory, the nonperturbative
superpotential is 
\beq
\label{wsu2dual}
W_{dyn}~=~\left( {\rm Pf} V\over 
{\bar{\Lambda}}_{2 L} \right)^{1/3}~.
\eeq 
This superpotential
also exists in supersymmetric QCD with $N_f > N_c$; upon adding mass
terms and integrating out the flavors, it gives rise to the usual 
superpotential induced by
gaugino condensation.\footnote{
We note an additional subtlety here: the scale of the $SU(2)$ dual 
theory below the confining scale of $SU(5)$ is field dependent,
${\bar{\Lambda}}_{2 L} \sim B$; for details, see \cite{PST1, PST2}.}

Thus, following the intricate  renormalization group 
flow, we arrive at a low-energy description in
terms of chiral superfields only. 
The moduli of this low-energy theory are  
$K_a^i, (r^2)^{ab}, (N\cdot r)^a_b, (N^2)_{ab}, B$, and $ \bar{B}$. 
The superpotential of this effective
theory is the sum of (\ref{wdual53B}) and (\ref{wsu2dual}).
We now turn on the tree level
superpotential (\ref{wtree53}), written in terms of the appropriate variables in
the confined low-energy description,
\beq
\label{wtree53dualmap}
W_{tree}~=~  \tilde{\lambda}_i^a ~ R_a \cdot Q \cdot L^i ~+
 ~ \tilde{\alpha}_{a b} ~ R_c
\cdot R_d \cdot R_e ~\epsilon^{a b c d e} ~=~ \tilde{\lambda}_i^a ~K_a^i ~+~
\tilde{\alpha}_{ab} ~(r^2)^{ab}~ .
\eeq
Note that the trilinear couplings in the tree level superpotential
are mapped, by the strong coupling dynamics,
 into linear terms in the low-energy superpotential.
We can now analyze the $F$ term conditions in the effective theory 
that follow from its superpotential:
\beq
\label{wfinal53}
W~=~  (N \cdot r)^a_a ~+~ 
A \left( N^2 \cdot K^3 - B \bar{B} - \bar{\Lambda}_{5}^{10} 
\right) ~ + ~\left( {\rm Pf} V\over 
{\bar{\Lambda}}_{2 L} \right)^{1/3}~ + \tilde{\lambda}_i^a ~K_a^i ~+~
\tilde{\alpha}_{ab} ~(r^2)^{ab}~.
\eeq
Extremizing with respect to $A$,  $K, r^2, N\cdot r, N^2, B$, and $\bar{B}$, we 
  find
that there is no extremum of the superpotential (\ref{wfinal53}), 
establishing thus that supersymmetry is broken 
(for more details, see \cite{PST2}).  Thus, as promised,  the   theory breaks 
 supersymmetry. 
Unfortunately,  we cannot say more
about the resulting vacuum.  
As mentioned above,  the  $SU(5) \times SU(2)$ dual description is  non-calculable.
 
We conclude this section with one comment.  The model discussed here, like many
others in this review are based on a  non-simple, product group gauge theory. 
In fact many of the  recently found theories 
with dynamical supersymmetry 
breaking  have product gauge groups \cite{DNNS, IT1, CRS96a, PST1,
PST2, CRSL, IT2, LRR, LT97}. This is in large part because the non-perturbative
behavior of such theories is  reasonably well understood---for example, 
as we saw above,   often dual theories can be constructed by applying Seiberg
duality to each  factor  of the product in turn \cite{PST1}. At the same time, 
as the  examples  here have shown, the interplay between the  various groups 
can lead to  interesting non-perturbative dynamics including supersymmetry
breaking.

\subsection{The $SU(5)$ model and related examples.} 

We end  this Section by discussing  some examples of models  where the 
scale of supersymmetry breaking and strong dynamics are comparable.   
The particular  models  we  study were  discovered long ago by  
Affleck, Dine, 
and Seiberg \cite{ADS4}, \cite{ADS5}.  Of course in many examples 
discussed in the previous sections  there is generically  (i.e. in the 
absence of a small coupling) no separation
between  the various scales.  
Some of our comments will be applicable to them as
well. 

The theory we look at is a ``one generation" $SU(5)$ model.
It has an 
   $SU(5)$  gauge symmetry  with a
single antisymmetric tensor,   $A \sim \Yasymm$,
 and antifundamental, $\bar{Q} \sim \overline{\Yfund}$, matter
representation.  Affleck,  Dine, and  Seiberg
argued that this theory  probably breaks supersymmetry. 

One can show that  the theory
has two anomaly  free global symmetries, $U(1)_{\cal A} \times U(1)_R$, under
which the superfields transform  as: $A \sim (-1, 1)$, $\bar{Q} \sim (3, -9)$.  These
global symmetries will  play an important role in the subsequent discussion. 
No   holomorphic gauge invariant 
can be constructed from the matter fields in this theory.  From this it follows
that  the theory has  no classical flat directions. It  also follows that  no
superpotential  is allowed by the gauge symmetries. 

The reasoning leading to the conclusion
that supersymmetry is  broken in this model goes as follows. If the global
symmetry of the model is unbroken there should be massless fermions in the spectrum
to saturate 't Hooft's anomaly matching conditions. The authors of 
\cite{ADS4} performed a search for solutions of the anomaly conditions 
and found that the simplest solutions were extremely complicated. To illustrate 
this point, we give  
one of the
simplest solutions of the  
anomaly matching conditions, tr$R = -26$, 
tr${\cal A} = 5$, tr${\cal A}^3 = 125$, tr$R^3 = -4976$, 
tr$R {\cal A}^2 = -450$, tr${\cal A} R^2 = 1500$.
The minimal solution \cite{ADS4} requires the existence of
 five massless Weyl fermions with charges 
$(-5,26)$, $(5,-20)$, $(5,-24)$, $(0,1)$, and $(0,-9)$ 
under the global 
$U(1)_{\cal A} \times U(1)_R$ symmetry (this is to be contrasted 
with the solution of the nonsupersymmetric 
version of the model, where a single massless fermion saturates
 't Hooft  anomaly matching \cite{Dimop}). 
The difficulty in satisfying anomaly matching leads to the conclusion  
  that some (or all) of the global symmetry is broken.
Now we can apply the general reasoning (Section 4.4) that
 if a global symmetry is broken in a theory without classical flat directions,
then supersymmetry is broken. One thus 
concludes 
that supersymmetry is broken in this theory.
Additional arguments  that
supersymmetry is broken, 
based on considering 
correlators in instanton backgrounds, appear in \cite{MV}. The scale of
supersymmetry
breaking is, presumably, of the order of the strong coupling scale of 
$SU(5)$, $\Lambda_5$, and
the massless spectrum should include a goldstino and Goldstone 
boson(s) for the 
broken global symmetry.

There exists a whole class of models, whose low-energy dynamics 
reduces to that
of the $SU(5)$ model. These are the  $SU(2 k + 1)$ theories with matter
representations  $A \sim \Yasymm$ and $\bar{Q}^i \sim \overline{\Yfund}$, with
$i = 1,...,2 k - 3$. These theories have classically flat directions, parametrized
by the gauge invariant holomorphic polynomials $X^{ij} = A \cdot \bar{Q}^i \cdot
\bar{Q}^j$. By studying the D-term equations, it is easy to see that along a 
generic flat direction, rank$\langle X^{ij}\rangle = 2 k - 4$, 
the gauge symmetry is
broken to an $SU(5)$ with a
$\Yasymm + \overline{\Yfund}$ matter representation \cite{ADS2}. 
This theory breaks 
supersymmetry
at a scale $\Lambda_{SU(5)}^{13} \sim 
\Lambda_{SU(2k+1)}^{4k+5}/X^{(4k-8)/3}$. 
The potential of the theory is, presumably,
proportional to $\Lambda_{SU(5)}^4$, 
and the theory has a runaway vacuum. The 
runaway behavior 
can be avoided if tree level terms  are added to the superpotential,
$W_{tree} = \lambda_{ij} X^{ij}$, to lift the classically flat directions.
The theory then  has a stable supersymmetry breaking vacuum.  

Another theory with very similar behavior to that of the 
$SU(5)$ ``one-generation" model is the
$SO(10)$ theory with a single spinor representation \cite{ADS5}.

It is worth mentioning that the Witten index for 
the $SU(5)$ and $SO(10)$ theories
can be calculated and vanishes,  consistent with supersymmetry breaking
as we discussed in Section 3.2. 
The basic idea is to add extra vectorlike flavors (e.g.  for the $SU(5)$ theory
pairs of  $\Yfund$ and $\overline{\Yfund}$) \cite{H1}.  The resulting theory 
now has D-flat directions. One  can add small mass terms  for the 
extra vector like flavors and analyze the 
low-energy dynamics in an effective supersymmetric field theory \cite{H1}, 
\cite{PT1}, \cite{Pouliot}, \cite{PS}.  One
finds that  supersymmetry is broken and thus that the Witten 
index is zero.  On
increasing  the masses, the extra heavy flavors decouple
 but  the Witten index stays  
unchanged.  
These theories 
with extra light flavors are interesting in their own right as 
examples of supersymmetry breaking. 
For example,  the 
 $SU(5)$ theory with two  extra pairs of  $\Yfund$ and $\overline{\Yfund}$
is a completely calculable model \cite{PT1}. 
The resulting theory is very similar to the 3-2 model: the gauge symmetry is 
totally broken, there is a nonperturbative
superpotential due to instanton effects, and for appropriately 
chosen parameters, the
vacuum occurs for large expectation values, where the whole spectrum is 
under perturbative control. 

\section{Phenomenological applications: gauge mediated supersymmetry
breaking.} 

Finally  we end this review by  briefly discussing 
 the application of 
dynamical supersymmetry breaking  to the construction
of phenomenological models of supersymmetry breaking.
For a  detailed review, a complete list of 
references, and discussion of phenomenological
signatures, we recommend  \cite{DDGR}.

As we discussed in the Introduction, 
supersymmetric extensions of 
the standard model offer an attractive solution to the hierarchy
problem. In  order to  explain the
hierarchy between the electroweak and Planck scales, 
supersymmetry has to break dynamically and generate 
the electroweak scale $m_W \sim  10^{-17} M_{Planck}$. 
Achieving dynamical supersymmetry breaking requires the addition of 
a new sector of the theory---the 
supersymmetry breaking sector. Some of the models we discussed in the 
previous sections  could play the roles of such a sector. 

In order to  generate masses for the
scalar partners of the quarks and leptons and the fermion partners
of the gauge bosons (and the soft parameters),
the breaking of supersymmetry has to be communicated to the standard model.
At present, theoretically speaking,  there exist two  main candidates for 
this messenger interaction. 

An obvious candidate for such a messenger interaction is supergravity. Until
recently, theories where supergravity is the messenger of supersymmetry
breaking were the most studied ones. There are good reasons for  
this: since
gravity is an universal interaction, once supersymmetry is broken in any sector
of the theory, it is automatically transmitted to all other sectors, generating
soft masses to the scalar superpartners of the quarks and leptons.
The soft masses are of order
\beq
\label{softsugra}
m_{soft} ~\sim ~{M_{SUSY}^2\over M_{Planck}} \sim 10^{2-3} {\rm GeV}~,
\eeq
where $M_{SUSY}$ is the supersymmetry breaking scale.
The above equation can be derived based on dimensional grounds: the soft
masses have to vanish in the limit $M_{Planck} \rightarrow \infty$, while 
the power
of $M_{Planck}$ follows from the fact that the communication is a tree-level
effect in the supergravity lagrangian. The requirement that the 
 the soft mass parameters are of order the electroweak scale follows from
phenomenological and naturalness considerations.
From eq.~(\ref{softsugra}), 
we can deduce that the scale of the supersymmetry breaking 
in supergravity mediated  
models is of order $M_{SUSY} \sim 10^{10-12}$ GeV. Thus, 
the supersymmetry breaking scale is rather high, beyond direct 
experimental reach. 
We will not discuss here the pros and cons of supergravity
mediated models of supersymmetry breaking, but  only mention an important
drawback:   the communication of supersymmetry breaking involves
dynamics at scales of order $M_{Planck}$, which is not  well understood at
present (many other shortcomings, such as generically large 
flavor changing neutral currents, can be related
 to this fact). 

An economical alternative to gravity,  as the messenger interaction, 
are the gauge interactions of the standard model. This scenario,
called gauge mediated supersymmetry breaking, has received considerable 
attention recently. 

In their simplest incarnation, gauge mediated models postulate the 
existence of new particles with standard model 
charges---the messenger
quarks and leptons. These messenger particles are heavy, with mass
of order $M_{mess}$. They  interact
with the supersymmetry breaking sector and  thereby acquire 
supersymmetry breaking mass splittings of order  $\Delta M_{mess}$.
Since they carry standard model gauge quantum numbers, the 
supersymmetry breaking mass splittings are transmitted to the standard
model squarks, sleptons, and gauginos at the loop level. Typically, a
soft mass parameter is of order
\beq
\label{msoftgauge}
m_{soft} ~\sim~{g^2 \over 16 \pi^2}~ \Delta M_{mess} ~\sim ~10^{2-3} {\rm GeV}~,
\eeq
where $g$ is a standard model gauge coupling and $1/16\pi^2$ is the standard
one-loop suppression factor. We see that the scale of the messengers is 
a lot smaller than the relevant scale
 in supergravity models: $\Delta M_{mess} \sim 10^{4-5}$ GeV.
A plausible possibility is that the scale of supersymmetry breaking is of
the same order as the scale of mass splittings of
 the messenger supermultiplets,
i.e. $M_{SUSY} \sim \Delta M_{mess}$.
We see then,
 that gauge mediated models of supersymmetry breaking  could involve
physics at scales much smaller than the scales in supergravity; one 
also makes no use of the ill-understood dynamics   at the Planck scale. 
The lower scale offers hope that the supersymmetry breaking dynamics 
may be amenable to direct experimental studies in a foreseeable future.
We saw in our discussion in Section 4.2 that the breaking of  global 
supersymmetry gives rise to a goldstino. In the presence of gravity
the goldstino is ``eaten" by the gravitino. Because of 
the low-scale of supersymmetry breaking in gauge mediated models the 
gravitino is often the lightest $R$ charged particle. This  can give rise 
to  distinct  experimental  signatures. 

Historically, gauge mediated models provided 
the first phenomenological framework
of supersymmetry breaking. After the advent of supergravity (in the early 1980's)
they were abandoned, mostly because of the alluring simplicity of supergravity
models (with the almost automatic generation of all soft parameters
at tree level), and  also 
because supersymmetric gauge dynamics was not well understood at the time.

Gauge mediated supersymmetry breaking was resurrected
in 1994, when the first phenomenologically viable 
model  was built \cite{DNS}. This model has served to refocus attention
on the possibility of gauge mediation  and provided an important 
``existence proof". 
But it has some drawbacks---one of them being its rather
complicated structure.  
The supersymmetry breaking sector in this model
 uses the (3,2)  model discussed in Section 5.1.1. 
The  anomaly free 
unbroken $U(1)_Y$ global symmetry is gauged; 
it is called the ``messenger $U(1)$" interaction. 
The messenger $U(1)$ transmits supersymmetry breaking to some
other fields, which in turn give a supersymmetry breaking expectation value to
a gauge singlet field (the ``messenger singlet"). 
The messenger singlet finally interacts with 
the messenger quarks and leptons and 
gives them the desired supersymmetry breaking mass splitting.
One reason for the complicated nature of this model is that at the time it was
constructed only a few theories of dynamical supersymmetry breaking
were known.

As we have seen in this review, recent studies of dynamical supersymmetry 
breaking have brought many new theories exhibiting this phenomenon
to light.   
These have proved instrumental in constructing new 
examples of gauge mediated models. 
In particular, one idea that has been explored  is to construct models 
where the unification of
the supersymmetry breaking and messenger sectors
(thereby getting rid  of the messenger $U(1)$ gauge interaction) is
achieved.  
 The idea is to identify the standard model gauge group with the unbroken
 global symmetry group of the supersymmetry breaking sector, 
which had to be large enough to accommodate the 
whole $SU(3)\times SU(2) \times U(1)$. 
The  states in the supersymmetry
breaking sector now  also carry standard model gauge quantum numbers.
Thus the messengers are identified with fields in the supersymmetry breaking
sector; their supersymmetry breaking mass splittings are transmitted to the
standard model squarks, sleptons, and gauginos at the loop level.

The first ``direct gauge mediation" models 
that were constructed 
used the $SU(N) \times SU(N-k)$, $N$-odd, $k = 1,2$ theories 
of \cite{PST1}, \cite{PST2}. For appropriate choices of $N$ these
theories have a ground state with broken supersymmetry and a 
sufficiently large global symmetry group to allow for embedding the
standard model gauge group in it \cite{PT2}, \cite{AHMMR}.
We also mention here that 
the ``plateau" models of ref.~\cite{H2}, \cite{DDGR}
have proven useful in constructing phenomenological models of
supersymmetry breaking. 
For a discussion
of the status  of the various  models mentioned here,
and other recent developments in the field, we refer the reader to \cite{GR}.

\section{Acknowledgments.}

We thank   Hsin-Chia Cheng, Ken Intriligator, Yael Shadmi, and Witek Skiba for
discussions and comments. 
E.P. is supported by DOE contract no.~DOE-FG03-97ER40506.
S.P.T.  is supported by the Fermi National
Accelerator Laboratory, which is operated by Universities Research
Association, Inc., under contract no.~DE-AC02-76CHO3000.

\nc{\ib}[3]{ {\em ibid. }{\bf #1} (19#2) #3}
\nc{\np}[3]{ {\em Nucl.\ Phys. }{\bf #1} (19#2) #3}
\nc{\pl}[3]{ {\em Phys.\ Lett. }{\bf #1} (19#2) #3}
\nc{\pr}[3]{ {\em Phys.\ Rev. }{\bf #1} (19#2) #3}
\nc{\prep}[3]{ {\em Phys.\ Rep. }{\bf #1} (19#2) #3}
\nc{\prl}[3]{ {\em Phys.\ Rev.\ Lett. }{\bf #1} (19#2) #3}


\begin{thebibliography}{99}

\baselineskip=14pt

\bibitem{wittendsb}E. Witten, \np{B188}{81}{513}.

\bibitem{S2}N. Seiberg, \pl{B318}{93}{469}.

\bibitem{S1}N. Seiberg, \pr{D49}{94}{6857}.

\bibitem{S3}N. Seiberg, \np{B435}{95}{129}.

\bibitem{IS}K. Intriligator and N. Seiberg, {\em Nucl. Phys. Proc. Suppl.} {\bf 45BC}
 (1996) 1.
 
\bibitem{Peskin}M. Peskin, Lectures at TASI-96, hep-th/9702094.
 
\bibitem{Shifman}M. Shifman, {\em Progr. Part. Nucl. Phys.} {\bf 39} (1997) 1. 

\bibitem{ADS1}I. Affleck, M. Dine, and N. Seiberg, \prl{51}{83}{1026};

I. Affleck, M. Dine, and N. Seiberg,  \np{B241}{84}{493}.

\bibitem{ADS2}I. Affleck, M. Dine, and N. Seiberg, \np{B256}{85}{557}.

\bibitem{ADS3}I. Affleck, M. Dine, and N. Seiberg, \prl{52}{84}{1677}.

\bibitem{ADS4}I. Affleck, M. Dine, and N. Seiberg, \pl{B137}{84}{187}.

\bibitem{ADS5}I. Affleck, M. Dine, and N. Seiberg, \pl{B140}{84}{59}.


\bibitem{ISS}K. Intriligator, N. Seiberg, and S. Shenker, \pl{B342}{95}{152}.

\bibitem{DNNS}M. Dine, A.E. Nelson, Y. Nir, and Y. Shirman, \pr{D53}{96}{2658}.

\bibitem{AN1}A.E. Nelson, \pl{B369}{96}{277}.

\bibitem{H1}H. Murayama, \pl{B355}{95}{187}.

\bibitem{PT1}E. Poppitz and S.P. Trivedi, \pl{B365}{96}{125}.

\bibitem{CSS1}C. Csaki, M. Schmaltz, and W. Skiba, \np{B387}{97}{128}.

\bibitem{IT1}K. Intriligator and S. Thomas, \np{B472}{96}{121}.

\bibitem{IY}K.-I. Izawa and T. Yanagida, 
{\em Progr. Theor. Phys.} {\bf 95} (1996) 829.

\bibitem{CRS96a}C. Csaki, L. Randall, and W. Skiba, \np{B479}{96}{65}.

\bibitem{PST1}E. Poppitz, Y. Shadmi, and S.P. Trivedi, \np{B480}{96}{125}.

\bibitem{PST2}E. Poppitz, Y. Shadmi, and S.P. Trivedi, \pl{B388}{96}{561}.

\bibitem{CRSL}C. Csaki, L. Randall, W. Skiba, and R.G. Leigh, \pl{B387}{96}{791}.

\bibitem{IT2}K. Intriligator and S. Thomas, hep-th/9608046.

\bibitem{LRR}R.G. Leigh, L. Randall, and R. Rattazzi, \np{B501}{97}{375}.

\bibitem{CSS2}C. Csaki, M. Schmaltz, and W. Skiba, \pr{D55}{97}{7840}.

\bibitem{LT97}M. Luty and J. Terning, hep-ph/9709306.


\bibitem{WS}W. Skiba, {\em Mod. Phys. Lett} {\bf A12} (1997)  737.

\bibitem{AN}A.E. Nelson, in Proc. International 
Conference on Supersymmetries in Physics
(SUSY 97), Philadelphia, PA, USA, May 1997, hep-ph/9707442.

\bibitem{EP}E. Poppitz, hep-ph/9710274, {\it Int. J. Mod. Phys.} {\bf A}, in press.

\bibitem{T1}S. Thomas, in Proc. International
 Workshop on Perspectives of Strong Coupling 
Gauge Dynamics (SCGT 96), Nagoya, Japan, Nov. 1996, hep-th/9801007.


\bibitem{AKMRV}D. Amati, K. Konishi, Y. Meurice, G.C. Rossi, 
and G. Veneziano, \prep{162}{88}{169}.

\bibitem{nonrenorm}M.T. Grisaru, W. Siegel, and M.M. Ro\v cek, \np{B159}{79}{429}.

\bibitem{APS}P.C. Argyres, M.R. Plesser, and N. Seiberg, \np{B471}{96}{159}.

\bibitem{GK}for a review, see: A. Giveon and D. Kutasov, hep-th/9802067.

\bibitem{LT}M. Luty and W. Taylor, \pr{D53}{96}{3399}.

\bibitem{WB}J. Wess and J. Bagger, {\it Supersymmetry and Supergravity},
(Princeton University Press, Princeton, NJ, 1993).

\bibitem{PR1}E. Poppitz and L. Randall, \pl{B389}{96}{280}.

\bibitem{svh}P. Salomonson and J.W. van Holten, \np{B196}{82}{503}.

\bibitem{wittenindex}E. Witten, \np{B202}{82}{253}.

\bibitem{wittennewgauge}E. Witten, hep-th/9712028.

\bibitem{Nilles}H.-P. Nilles, \pr{110}{84}{1}.

\bibitem{Huq}M. Huq, \pr{D14}{76}{3548}.

\bibitem{Shirman}Y. Shirman, \pl{B389}{96}{287}.

\bibitem{H2}H. Murayama, \prl{79}{97}{18}.

\bibitem{DDGR}S. Dimopoulos, G. Dvali, G.F. Giudice, and R. Rattazzi, 
\np{B510}{97}{12}.

\bibitem{AHM1}N. Arkani-Hamed and H. Murayama, hep-th/9705189.

\bibitem{GR}G.F. Giudice and R. Rattazzi, hep-ph/9801271.


\bibitem{DSW}M. Dine, N. Seiberg, and E. Witten, \np{B289}{87}{589}.


\bibitem{NS}A.E. Nelson and N. Seiberg, \np{B426}{94}{46}.

\bibitem{BW}J. Bagger and E. Witten, \pl{B115}{82}{202}.

\bibitem{PR}E. Poppitz and L. Randall, \pl{B336}{94}{402}.

\bibitem{BPR}J.A. Bagger, E. Poppitz, and L. Randall, \np{B426}{94}{3}.

\bibitem{HKLR}N.J. Hitchin, A. Karlhede, U. Lindstr\" om, and M. Ro\v cek, 
{\it Comm. Math. Phys.} {\bf 108} (1987) 535.

\bibitem{Veldhuis}T.A. ter Veldhuis, \pl{B367}{96}{157};

T.A. ter Veldhuis, hep-th/9802125.

\bibitem{DMS}G. Dotti, A. Manohar, and W. Skiba, in preparation.

\bibitem{Berkooz}M. Berkooz, \np{B452}{95}{513}.

\bibitem{Kutasov}D. Kutasov, \pl{B351}{95}{230};

D. Kutasov and A. Schwimmer, \pl{B345}{95}{315};

D. Kutasov, A. Schwimmer, and N. Seiberg, \np{B459}{96}{455}.

\bibitem{PT2}E. Poppitz and S.P. Trivedi, \pr{D55}{97}{5508}.

\bibitem{AHMMR}N. Arkani-Hamed, J. March-Russell, and H. Murayama,
\np{B509}{97}{3}.


\bibitem{WittenIH}E. Witten, \pl{B105}{81}{267}.

\bibitem{Dimop}S. Dimopoulos, S. Raby, and L. Susskind, \np{B173}{80}{208}.

\bibitem{MV}Y. Meurice and G. Veneziano, \pl{B141}{84}{69}.

\bibitem{Pouliot}P. Pouliot, \pl{B367}{96}{151}.

\bibitem{PS}P. Pouliot and M. Strassler, \pl{B375}{96}{175}.

\bibitem{DNS}M. Dine, A.E. Nelson, and Y. Shirman, \pr{D51}{95}{1362}.

\end{thebibliography}
\end{document}